\documentclass{aa}
\usepackage{psfig}
\begin{document}
\def\point{$\cdot$ \hspace{0.1cm} $\cdot$ \hspace{0.1cm} $\cdot$}
   \thesaurus{03     
              (04.19.1;  
               11.01.2;  
               11.03.1;  
               11.17.2;  
               11.19.1;  
               13.25.2;)} 
   \title{The ROSAT Deep Survey}

    \subtitle{III. Optical spectral properties of X--ray sources in the Lockman Hole}

   \author{I. Lehmann\inst{1}
          \and
          G. Hasinger\inst{1}
          \and
          M. Schmidt\inst{2}
          \and
          J.E. Gunn\inst{3}
          \and
          D.P. Schneider\inst{4}
          \and
          R. Giacconi\inst{5}
          \and
          M. McCaughrean\inst{1}
          \and
          J. Tr\"umper\inst{6}
          \and
          G. Zamorani\inst{7,}\inst{8}
          }

   \offprints{I. Lehmann; ilehmann@aip.de}

   \institute{Astrophysikalisches Institut Potsdam (AIP),
              An der Sternwarte 16, D-14482 Potsdam, Germany
         \and
             California Institute of Technology, Pasadena, CA 91125, USA
         \and
             Princeton University Observatory, Peyton Hall, 
             Princeton, NJ 08540, USA
         \and
             Department of Astronomy and Astrophysics, The Pennsylvania
             State University, University Park, PA 16802, USA
         \and
             European Southern Observatory, Karl-Schwarzschild-Str. 1,
             D-85748 Garching bei M\"unchen, Germany
         \and
             Max-Planck-Institut f\"ur extraterrestrische Physik,
             Karl-Schwarzschild-Str. 2, D-85748 Garching bei M\"unchen,
             Germany
         \and
             Osservatorio Astronomico, Via Ranzani 1, I-40127, Bologna,
             Italy
         \and
            Istituto di Radioastronomia del CNR, via Gobetti 101,
             I-40129, Bologna, Italy
             }

   \date{Received September , 1999; accepted November 22, 1999}

   \maketitle

   \begin{abstract}
The ROSAT Deep Survey in the Lockman Hole contains a complete sample
of 50 X-ray sources with fluxes in the 0.5-2 keV band larger than
5.5 $\cdot$ 10$^{-15}$ erg cm$^{-2}$ s$^{-1}$. Previous work has provided
optical identification  of 46 of the 50 X-ray sources; over 75 \% of the 
sources are AGNs (Schmidt et al. \cite{Schm98}).

We present now the atlas of optical finding charts and the full
description of the spectra, including emission line 
properties of the optical counterparts, which are important for the object 
classification.

New optical/infrared observations of three of the four unidentified
sources show that one source is an AGN and two sources with an unusually 
large ratio of X-ray to optical flux have counterparts in the K'-band suggesting
that they are obscured AGNs.
Furthermore, we found evidence from radio emission that the 
remaining unidentified source is a powerful radio galaxy (AGN).
We thus obtain a 100~\% completeness.

During the course of our optical identification work, we obtained optical spectra of 83 field galaxies, of which 67 were narrow--emission line galaxies (NELG). We demonstrate that it is highly unlikely that a significant number of NELG are
physically associated with X--ray sources.

      \keywords{surveys -- galaxies: active -- galaxies: clusters: general --
                galaxies: Seyfert -- quasars: emission lines --
                X-rays: galaxies
               }
   \end{abstract}

%

\section{Introduction}

A reliable optical identification of faint X-ray sources from deep X-ray surveys
is the key to determine the contribution of the various types of X-ray
sources to the X-ray background (XRB), originally discovered more than three decades ago (Giacconi et al. \cite{Giac62}). 

The ROSAT Deep Survey (RDS) in the Lockman Hole includes a statistically well defined and complete sample of 50 X-ray sources with fluxes in the 0.5-2 keV band larger than 5.5 $\cdot$ 10$^{-15}$ erg cm$^{-2}$ s$^{-1}$ (Hasinger et al. \cite{Has98}, Paper I). The survey has been carried out with
the ROSAT satellite in the Lockman Hole (Hasinger et al. \cite{Has93}), a region
of extremely low galactic hydrogen column density (Lockman et al. \cite{Loc86}). The RDS is based on a 207 ksec ROSAT PSPC exposure, a 205 ksec HRI raster scan and a series of HRI exposures of a fixed area totaling 1112 ksec. The deep HRI images produced very accurate (2\arcsec) positions, which were crucial for the optical identification effort. 
The extraction of X-ray sources and their X-ray properties are described in Paper I.

The optical identification and photometry of the RDS sources have been published
in Paper II (Schmidt et al. \cite{Schm98}). Among the 50 X-ray sources we identified 
39 AGNs, 3 groups of galaxies, 1 galaxy and 3 galactic stars; four X-ray sources were not identified. This study found a much larger fraction of AGNs than did any previous X--ray survey (Boyle et al. \cite{Boy95a}, Georgantopoulos et al.
\cite{Geor96}, Bower et al. \cite{Bow96} and McHardy et al. \cite{Mc98}).

In this paper we present the optical finding charts of the RDS sources and their optical spectra as well as some new optical/near-infrared and radio identifications.  
The paper is organized as follows. In Sect. 2 we discuss the 
optical observations and the data reduction. In Sect. 3 we present 
the atlas of the optical finding charts, the spectra and the galaxy absorption lines of the optical counterparts of the X-ray sources. Sect. 4 contains 
a new optical identification as well as a possible identification of two X-ray sources with infrared sources and a probable identification of a radio galaxy.
On the basis of the optical spectra we have determined some emission line properties, which
are presented in Sect. 5. In Sect. 6 we discuss the fraction of narrow emission line galaxies (NELG) in the field near the RDS sources and discuss the possible misidentification of faint X--ray sources with the class of NELG in several deep X-ray surveys. A brief discussion of the results is given in Sect. 7.
Throughout the paper we use H$_{0} = 50$ km s$^{-1}$ Mpc$^{-1}$ and q$_{0} = 0.5$.

\section{Observations and Reductions}

\subsection{Optical imaging}

$R$-band images of the Lockman field were obtained with the 
Low Resolution Imaging Spectrometer (LRIS; Oke et al. \cite{Oke95}) on the 
Keck I telescope in April 1996 and on the Keck II telescope in April 1997 and 
March 1998. The image scale on the back-illuminated  2048 $\times$ 2048
Tektronics CCD is 0.215 arc--sec per pixel. The exposure times of the $R$-images were 300 sec
or 30--60 sec, when the $R$-image was used for the aquisition of a multi-slit
mask. The typical seeing was around $\approx$1\arcsec. The limiting
magitudes of the $R$-images range from 23.5 to 26.0 mag. The 6\arcmin~x 8\arcmin~field of view of the LRIS is sufficiently large that each image contains at least two RDS sources. The images were debiased and flattened using standard procedures. 
The detection and photometric analysis of the images was done using the SExtractor package (Bertin \& Arnouts \cite{Ber96}). The zero-point of the photometry was obtained by a cross-correlation with a digital version of the POSSII (Djorgovski priv. communication), which is within 0.2 mag of the standard instrumental calibration. The photometric error is around 0.5 mag in $R$.

For all sources we have in addition $V$-- and $I$--band images from Palomar CCD drift scans, which are described in Paper II.

\subsection{Optical spectroscopy}

The optical spectroscopy of the counterparts of most X-ray sources was
carried out in February, December 1995, April 1996, April 1997 and in March 1998 at the Keck I and II telescopes on Mauna Kea, Hawaii.
The majority of spectra were obtained using the LRIS instrument in the long-slit mode with a 0.7\arcsec or 1.0\arcsec~wide slit. The spectra of the objects 12A, 16A, 23A and 116B/D were taken through 1.5\arcsec~wide slits using LRIS multi-slit masks. The detector was the same as used for imaging (see above). 
A 300 lines  mm$^{-1}$ grating with a wavelength coverage of around 3800--8200 \AA~yields a spectral dispersion of 2.46~\AA~per pixel and a resolution of $\sim$ 10-15 \AA~(FWHM). A typical
integration was a single 1800 sec exposure in the long-slit mode or two exposures of 1800 sec in the multi-slit mode. 

Optical spectra for the relatively bright objects 8B, 9A, 20C, 25A, 27A, 28B 
and 52A were taken in February and December 1992 with the 4-Shooter spectrograph (Gunn et al. \cite{Gun87}) at the 5-m Hale telescope. The detector is
a 800 $\times$ 800 Texas Instrument CCD. Using a 
1.5 $\times$ 100\arcsec~slit and a 200 lines mm$^{-1}$ transmission
grating the spectra cover a wavelength range from 4500-9500~\AA~at a
resolution of 25 \AA. The data analysis was performed using
the procedure given in Schneider et al. (\cite{Schn94}).

An optical spectrum of the bright star 62A was obtained in March 1998 
with the Boller \& Chivens Cass Twin spectrograph at the 3.5-m telescope 
on Calar Alto. A 1.5\arcsec~wide slit and the 600 lines mm$^{-1}$ transmission
grating (T13) was used resulting in a wavelength coverage of 3500-5500 \AA~and
a resolution of about 4.4 \AA~in the blue channel. The red channel
was not available due to technical problems with the CCD. The detector is
a 800 $\times$ 2000 SITe CCD. 

The Keck spectra and the Calar Alto spectrum were processed with the software
package MIDAS. The raw data frames were bias-subtracted and flatfield corrected.
Following a night sky-subtraction, where we selected a small region of constant background 
light nearby the sources, the extraction of the one-dimensional
spectra was done using the optimal extraction algorithm described by Horne
(\cite{Hor86}). The wavelength scale was defined by fitting third order polynomials
to the lines of a He-Ar or Hg-Kr calibration spectrum. The relative flux
calibration was obtained using secondary standard stars for spectrophotometry
(Oke \& Gunn \cite{Oke83}). The atmospheric correction function for the 
broad molecular absorption bands was derived from the spectra of the standard
stars and applied to all spectra. However, for some spectra (see eg. 35A and 37A in
Fig. \ref{spec1}) the atmospheric correction did not work reliably. 

\begin{figure*}
     \begin{minipage}{177mm}
\psfig{figure=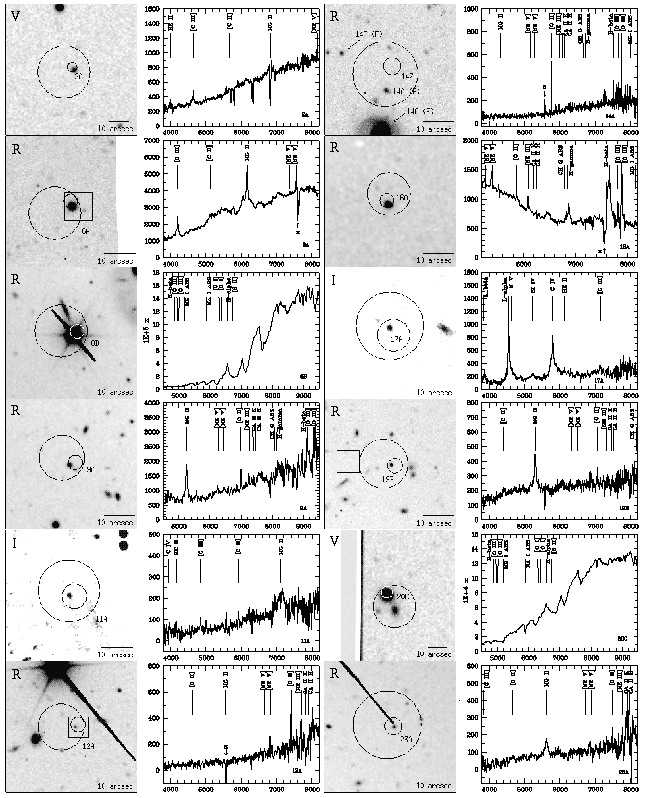,bbllx=66pt,bblly=55pt,bburx=556pt,bbury=664pt,width=177mm}
     \end{minipage}
     \caption[]{$R$-band/$V$-band and $I$-band images and low-resolution spectra
     of the RDS sources. North is up, and east to the left. The small and large circles
     show the ROSAT HRI and PSPC error circles (80 \% error radius). The 2$\sigma$ error
     box indicates a VLA 20 cm detection. The wavelength is given in \AA ngstrom, the flux scale 
     is normalized that one count corresponds to an AB magnitude 28 
     (f$_{\nu}=2.29 \cdot 10^{-31}$ erg s$^{-1}$ cm$^{-2}$ Hz$^{-1}$).
     With ''x'' a region ($\sim$ 7600 \AA) of a bad atmospheric correction 
     is marked.  Residuals of night sky emission lines are indicated with ''s''. }
     \label{spec1}
\end{figure*}
%
\addtocounter{figure}{-1}
%
\begin{figure*}
     \begin{minipage}{177mm}
\psfig{figure=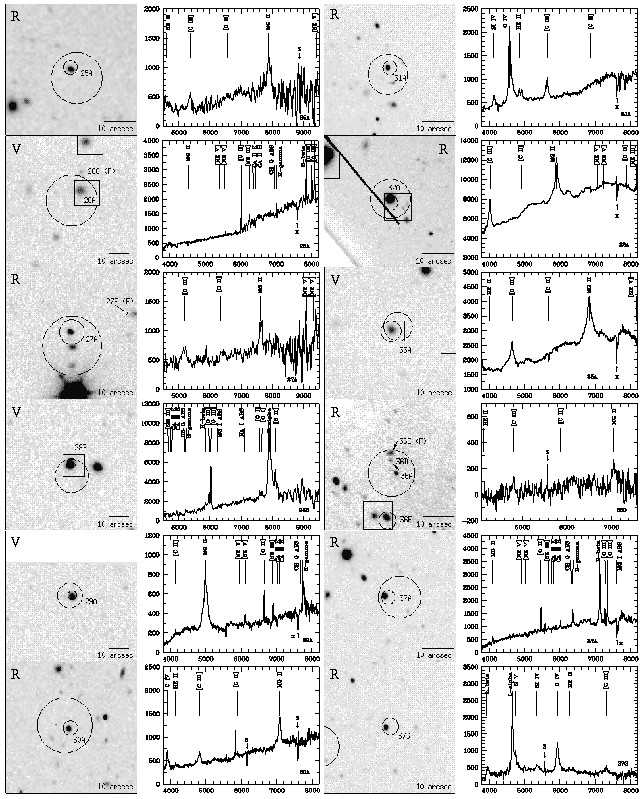,bbllx=66pt,bblly=55pt,bburx=556pt,bbury=664pt,width=177mm}
     \end{minipage}
     \caption[]{--Continued.}
     \label{spec2}
\end{figure*}
%
\addtocounter{figure}{-1}
%
\begin{figure*}
     \begin{minipage}{177mm}
\psfig{figure=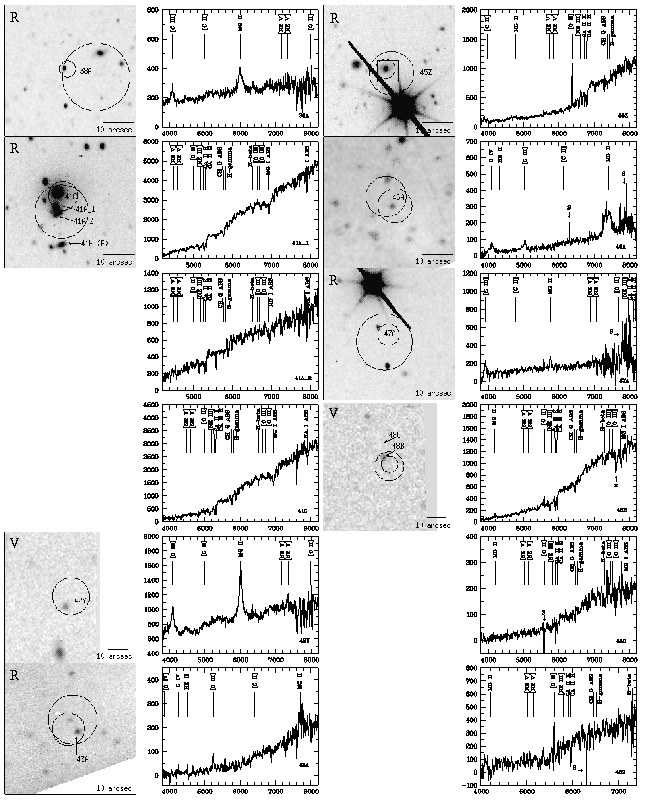,bbllx=66pt,bblly=55pt,bburx=556pt,bbury=664pt,width=177mm}
     \end{minipage}
     \caption[]{--Continued.}
     \label{spec3}
\end{figure*}
%
\addtocounter{figure}{-1}
%
\begin{figure*}
     \begin{minipage}{177mm}
 \psfig{figure=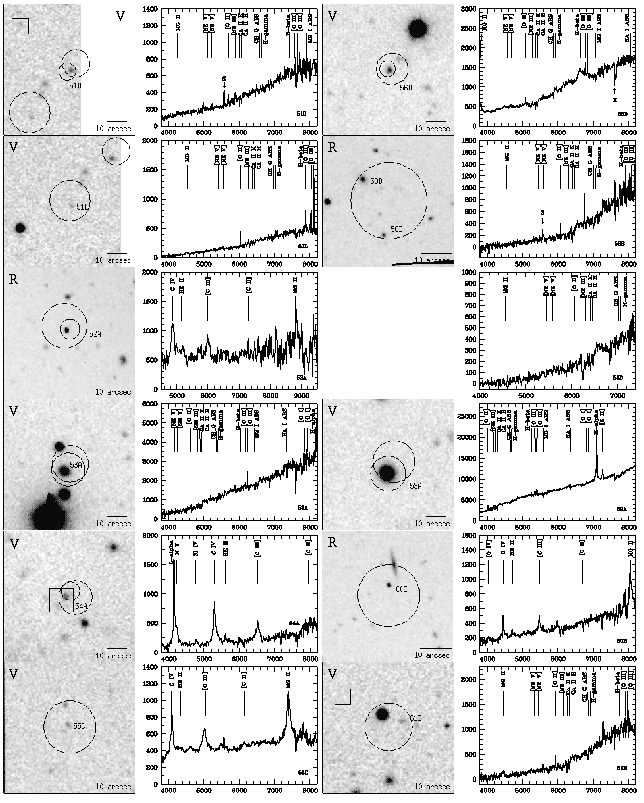,bbllx=66pt,bblly=55pt,bburx=556pt,bbury=664pt,width=177mm}
     \end{minipage}
     \caption[]{--Continued.}
     \label{spec4}
\end{figure*}
%

\addtocounter{figure}{-1}
%
\begin{figure*}
     \begin{minipage}{177mm}
 \psfig{figure=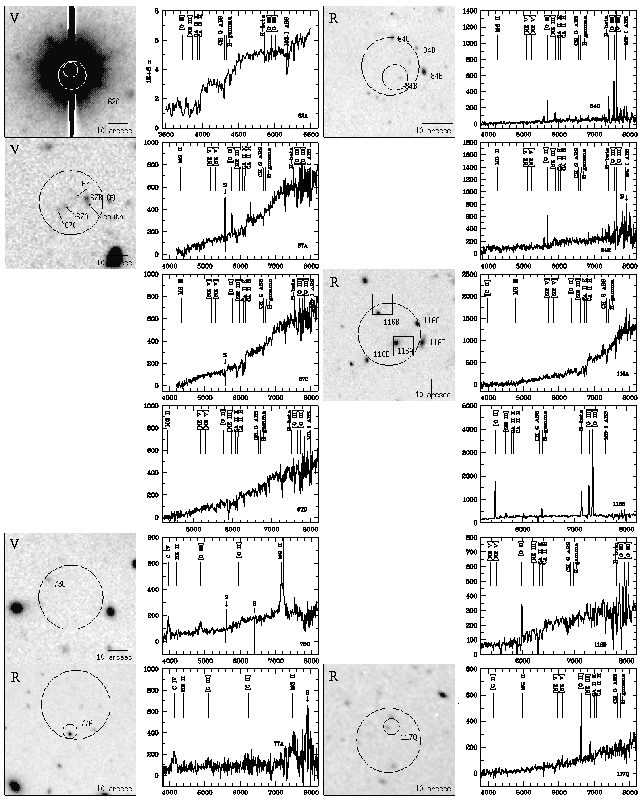,bbllx=66pt,bblly=55pt,bburx=556pt,bbury=664pt,width=177mm}
     \end{minipage}
     \caption[]{--Continued.}
     \label{spec5}
\end{figure*}
%

\section{Atlas of optical finding charts and spectra}

The finding charts for all X-ray sources are ordered by increasing
X-ray source number (cf. Fig. \ref{spec1}). For most sources the optical
 images are taken from Keck $R$-images, except for the sources  2, 11, 17, 20, 28, 29, 35, 42, 48, 51, 53, 54, 55, 56, 59, 61, 62, 67 and 73, for most of which the $V$-images from the Palomar drift scans (see Paper II) are shown.
The $V$-images of the sources 11 and 17 are not deep enough to show the suggested optical counterparts. Therefore we present the finding charts of both sources using deep raw $I$-images, kindly provided by G. Luppino (priv. communication), taken with the 8K CCD Mosaic Camera (Metzger et al. \cite{Me95}) at the University of Hawaii 2.2m telescope.

The image scale is given by the horizontal bar.
The two circles drawn on the images correspond to the 80 \% PSPC
and HRI error circles, which are derived from the 207 ksec PSPC image, the 205 ksec HRI raster scan and the 1112 ksec HRI images. 
The HRI error circle is always smaller than the PSPC circle due to the 
higher spatial resolution of the ROSAT HRI detector (5\arcsec) in comparison
with 20\arcsec~of the PSPC detector.

The HRI error circles of the X-ray sources  35, 36, 43, 46, 48, 53, 56, 59 and 62 were taken from the HRI raster scan exposures. The sources 37 and 51 are confusion resolved by the HRI exposure into the object pairs 37 A$+$G and 51 D$+$L, whereas the PSPC image shows only a single source in both cases. The HRI detection of the PSPC sources is
extremely important to identify the optical counterpart. We found in nearly all
cases only one optical object within the HRI error circle. PSPC data only were available for the sources 26, 36, 42, 55, 58, 60, 61, 67, 73 
and 116. In these cases the optical identification is much more difficult, because several 
objects frequently lie within the PSPC error (cf. 36 in Fig. \ref{spec1}). The optical counterparts of the X--ray sources are indicated by name in Fig. \ref{spec1}. For a number of field galaxies near the X--ray sources we have obtained optical spectra during the identification process. These objects are labeled with ''(F)''. 
 
A radio survey in the Lockman Hole has recently been carried out with the Very Large Array (VLA) at a wavelength of 20~cm (De Ruiter et al. \cite{deRui97}). About 149 radio sources were detected down to a limiting flux density of $\sim$120 $\mu$Jy. The cross correlation with the RDS X--ray source list has yielded 9 radio/X--ray associations (6, 12, 26, 28, 32, 35, 45, 54, and 116). VLA detections are illustrated by a square box in the finding charts. The positional errors were taken from De Ruiter et al. (\cite{deRui97}). For all sources in which the radio sources are within the X--ray error box, the association between the radio and the suggested optical counterpart is highly reliable (see Table~2 in
De Ruiter et al. \cite{deRui97}), with no positional offset larger than 2$\sigma$ in the combined positional errors.
 Taking into account the different X--ray and radio limits, the percentage of radio detections of the Lockman Hole X--ray sources (14\% at $\sim$0.2mJy)
is in good agreement with the fraction of $\sim9$\% for X--ray sources in the Cambridge-Cambridge ROSAT Serendipity Survey (Ciliegi et al.\cite{Ci95}) and the value of 8\% found in the Marano field (Zamorani et al. \cite{Zam99}). 
The VLA detection of source 116 is discussed in Sect. 4.3. 

In Fig. \ref{spec2} we present as well the optical spectra of the RDS sources, smoothed to the instrumental resolution (10/15 \AA~for the Keck long-slit/multi-slit spectra, 25 \AA~for Palomar spectra and 4.5 \AA~for the Calar Alto spectrum of
62A). The flux scale is normalized so that one count corresponds to an AB magnitude 28
 (f$_{\nu}=2.29 \cdot 10^{-31}$ erg s$^{-1}$ cm$^{-2}$ Hz$^{-1}$). 
The location of typical AGN emission lines (eg. Ly$\alpha$, C IV, C III], Mg II) and galaxy absorption lines (Ca H+K $\lambda$3934/3968, CH G $\lambda$4304, Mg I $\lambda$5175, Na I $\lambda$5890) are overplotted. The residuals of the atmospheric correction ($\sim$7600 \AA) are marked in some spectra (eg. for 6A and 37A).

Three of the brightest X--ray sources are foreground stars. The comparison of their optical spectra (cf. Fig. \ref{spec1}) with the library of stellar spectra from Jacobi et al. \cite{Jac84} reveals two M5 V stars (8B and 20C) and a K0 V star (62A).  

The high signal to noise Keck spectra are very important to identify the faint optical counterparts of the X--ray sources. Five ID classes are used to classify the 47 extragalactic X--ray sources.
The optical identification scheme (ID class a--e) is described more precisely in Paper II.
The optical counterparts of 33 X--ray sources have been identified with AGNs
of ID classes a-c, because they exhibit broad emission lines in their optical spectra. In this scheme the objects of ID class a show at least two of the following broad emission lines: Mg II, C III], C IV and Ly$\alpha$.
We have changed the ID class of source 27 from e (see Paper II) to a, because the source has been identified with the optical quasar 27A at $z=1.720$, which is inside the final HRI error circle. The object near the center of the PSPC circle is a normal galaxy without emission lines ($z=0.487$). 
The 7 objects of ID class b, at lower redshift, exhibit broad Mg~II emission lines and 3 of
them (56D, 37A and 9A) show as well broad emission lines of H${\beta}$. 
The 3 members of the ID class c have only broad emission lines of H${\beta}$ and/or H$\alpha$. Six X--ray sources were classified as AGNs of the ID class d. Their optical spectra contain no broad emission lines seen for the ID classes a-c, but four objects (12A, 51D, 51L and 117Q) exhibit [Ne V] emission lines indicating a high ionization potential, corresponding to soft X--ray emission typical for AGNs. The objects 26A and 45Z, both have X--ray luminosities in the 0.5--2.0 keV energy band above 10$^{43}$ erg s$^{-1}$, show relatively strong [Ne III] $\lambda3869$ emission lines, which are typically quite
strong in Seyfert galaxies and quasars (cf. Paper II.). 

Eight X--ray sources belong to the ID class~e. All the spectra of the proposed identifications
in this class do show neither broad lines nor neon forbidden lines. In this class there are three X-ray sources identified with groups or clusters of galaxies (41, 58 and 67), one AGN or group (48), one emission line galaxy (53), one radio galaxy (116), two very red objects (14, 84), detected in K$^{\prime}$ band and not detected in the optical bands. The first five of these sources are briefly described below, while new data and the suggested identifications for the remaining three sources are discussed in Sect. 4.

 The $R$-image of 41 reveals a number of faint galaxies at the position of the X--ray source, which is extended in the PSPC and in the HRI. The spectra of the cluster members 41C, 41A$-$1 and 41A$-$2 only show typical galaxy absorption lines at a redshift of 0.340. 
Five faint galaxies are detected inside the PSPC circle at the $V$-image of source 67. The objects 67A, 67B (south), 67C and 67D have the same redshift ($z=0.525$), whereas 67B is NELG at $z=0.317$. In the case of source 58 there are two galaxies 58B and 58D at a redshift of 0.629 near the inner edge of the PSPC circle. The spectrum of 58B shows a faint peak, labeled with ''S'', near the position of the [Ne V]$\lambda$3426 emission line, which is a result of a bad night sky subtraction around the prominent 5570 \AA~sky line. The source is classified as a group of galaxies (cf. Paper II). 

Source 48 has a faint galaxy 48B inside the HRI raster scan error circle. The galaxy 48C (at the same redshift) is located at the edge of the HRI error circle, but within the larger PSPC error circle. A further galaxy 48Z at the same redshift is at a distance of 85\arcsec~from the X-ray source. On the basis of the  high X-ray luminosity in the 0.5--2.0 keV energy band (log $L_{x}=43.3$) and the f$_{x}$/f$_{v}$ ratio, 48B was classified as an AGN, although this classification is uncertain (see Paper II), and the optical counterpart could also be a group of galaxies. 

Source 53 is the only X--ray source, which has been classified as an emission line galaxy. The optical spectrum of 53A exibits two narrow emission lines ([O III]$\lambda$5007 and H${\alpha}$). The X--ray luminosity in the 0.5--2.0 keV energy band of log $L_{x}=42.2$ appears to be rather high for a galaxy. The galaxy is located between two M--stars, which in principle could also contribute to the 
X--ray emission. The classification of source 53 is therefore somewhat uncertain and needs higher angular resolution as will be provided by Chandra. 

\subsection{Galaxy absorption lines and D(4000) break index}
In Table \ref{TAB1} we present the galaxy absorption lines of the optically identified RDS sources. The columns 1--3 contain the name of the optical/infrared counterpart, the classification  (AGN, Grp, star or galaxy) and the identification class (a--e). There are two pairs of sources with the same number (37 and 51). The object 37G is the ID of the source 814 and the object 51D is the ID of the source 504 in Table~1 of Paper II. In the case of a group or a cluster of galaxies there is only one entry for the brightest galaxy. Column 4 gives the redshift of the objects. The $R$-magnitude of the objects, which was taken from Paper II, is shown in column 5. The column 6 presents typical galaxy
absorption lines using the following code: A) Ca H+K $\lambda$3934/3968, B) CH G $\lambda$4304, C) Mg I $\lambda$5175, D) Na I $\lambda$5890 and ''no'' means that there are no Ca H+K absorption lines visible within the expected wavelength range. There is no entry for the RDS objects, where the mentioned absorption lines were not accessible due to a redshift larger than 1.1. 
The column 7 shows the 4000\AA~break index D(4000) and its 1$\sigma$ error, which is defined by Bruzual (\cite{Bru83}) as the ratio of the average flux density f$_{\nu}$ in the 4050--4250 \AA~and 3750--3950 \AA~bands in the rest frame. Due to the wavelength range of our spectra we were able to derive the D(4000) values for all objects with redshift between 0.01 and 0.93.

The fact that most of the spectra of the RDS AGNs in the appropriate
redshift range show one or more of the 'typical galaxies' absorption lines,
suggests that the non-thermal AGN emission is not dominant at these
wavelengths for these objects. 

%
   \begin{table}[t]
      \caption{Galaxy absorption lines of RDS objects.}
      \label{TAB1}
      \begin{flushleft}
     \begin{tabular}{l@{\hspace*{-1mm}}c@{\hspace*{-1mm}}c@{\hspace*{2mm}}c@{\hspace*{5mm}}c@{\hspace*{2mm}}c@{\hspace*{3mm}}c@{\hspace*{3mm}}c}
      \hline\noalign{\smallskip}
name & class & ID class  & $z$ & $R$ & Lines & D(4000) \\
\hspace{0.2cm}(1) & (2) & (3) & (4) & (5) & (6) & (7) \\\noalign{\smallskip}
\hline \noalign{\smallskip}
2A  & AGN & a & 1.437 &  20.1 &-&-\\
6A  & AGN & a & 1.204 &  18.4 &-&-\\
8B  & star&M5 V& -     &  14.1 &CD&-\\
9A  & AGN & b & 0.877 &  20.0 &AB&1.10$\pm$0.30\\
11A & AGN & a & 1.540 &  23.0:&-&-\\ \noalign{\smallskip}
12A & AGN & d & 0.990 &  22.9 &A&-\\ 
14Z & AGN:& e & -     &  25.0 &-&-\\
16A & AGN & c & 0.586 &  19.8 &A:&0.78$\pm$0.09\\
17A & AGN & a & 2.742 &  20.3 &-&-\\
19B & AGN & b & 0.894 &  21.8 &AB:&1.11$\pm$0.47\\ \noalign{\smallskip}
20C & star&M5 V& -     &  15.5 &CD&-\\
23A & AGN & b & 1.009 &  21.9 &A&-\\ 
25A & AGN & a & 1.816 &  20.6 &-&-\\
26A & AGN & d & 0.616 &  18.7 &AB&1.28$\pm$0.19\\
27A & AGN & a & 1.720 &  20.3 &-&-\\ \noalign{\smallskip}
28B & AGN & c & 0.205 &  18.2 &no&1.21$\pm$0.69\\
29A & AGN & b & 0.784 &  19.5 &A&1.18$\pm$0.18\\
30A & AGN & a & 1.527 &  21.5 &-&-\\
31A & AGN & a & 1.956 &  20.0 &-&-\\
32A & AGN & a & 1.113 &  18.1 &-&-\\ \noalign{\smallskip}
35A & AGN & a & 1.439 &  18.9 &-&-\\
36D & AGN & a & 1.524 &  22.5 &-&-\\
37A & AGN & b & 0.467 &  19.6 &A&1.11$\pm$0.10\\
37G & AGN & a & 2.832 &  20.5 &-&-\\
38A & AGN & a & 1.145 &  21.3 &-&-\\ \noalign{\smallskip}
41C & Grp & e & 0.340 &  17.9 &ABCD&2.02$\pm$0.45\\
42Y & AGN & a & 1.144 &  20.7 &-&-\\
43A & AGN & a & 1.750 &  23.0 &-&-\\
45Z & AGN & d & 0.711 &  21.1 &AB&1.62$\pm$0.34\\
46A & AGN & a & 1.640 &  22.6 &-&-\\ \noalign{\smallskip}
47A & AGN & a & 1.058 &  21.9 &no&-\\
48B & AGN:/Grp& e & 0.498 &  19.9 &ABC&1.87$\pm$0.56\\
51D & AGN & d & 0.528 &  20.2 &A&1.28$\pm$0.22\\
51L & AGN & d & 0.620 &  21.1 &A&1.39$\pm$0.27\\
52A & AGN & a & 2.144 &  20.4 &-&-\\ \noalign{\smallskip}
53A & Gal & e & 0.245 &  18.4 &ABCD&1.64$\pm$0.39\\
54A & AGN & a & 2.416 &  20.3 &-&-\\
55C & AGN & a & 1.643 &  21.4:&-&-\\
56D & AGN & b & 0.366 &  18.9 &ABC&1.34$\pm$0.15\\
58B & Grp & e & 0.629 &  21.4 &AB:&1.97$\pm$0.75\\ \noalign{\smallskip}
59A & AGN & c & 0.080 &  16.9 &ABCD&1.37$\pm$0.18\\
60B & AGN & a & 1.875 &  21.6 &-&-\\
61B & AGN & b & 0.592 &  20.8 &AB&1.82$\pm$0.60\\
62A & star& K0 V& -     &  11.0&ABC&- \\ 
67B(s) & Grp & e & 0.550 &  20.5:&AB&1.98$\pm$0.59\\ \noalign{\smallskip}
73C & AGN & a & 1.561 &  20.6 &-&-\\
77A & AGN & a & 1.676 &  21.7 &-&-\\
84Z & AGN:& e &   -   &$>$25.5&-&-\\
116A& AGN & e & 0.708 &  20.9 &AB&2.07$\pm$0.53\\
117Q& AGN & d & 0.780 &  22.8 &A&1.42$\pm$0.56\\
      \noalign{\smallskip}
      \hline
      \end{tabular}
      \end{flushleft}
      \begin{list}{}{}
       \item A) Ca H+K $\lambda$3934/3968, B) CH G $\lambda$4304, C) Mg I $\lambda$5175, D) Na I $\lambda$5890 in column 6.
       \end{list}
      \end{table}
\clearpage

Therefore, we have measured the break index D(4000) in order to have an estimate of the main stellar population in the host galaxies. 
About 63\% of the RDS AGNs for which the D(4000) index
could be derived from the optical spectra, have D(4000) $>$ 1 at more than 1 sigma level (see Fig. 2b).
The object 16A, whose spectrum shows a very blue continuum, is the only AGN with D(4000) $<$ 1, consistent with a power law continuum (f$_{\nu}=\nu^{\alpha}$ with $\alpha<0$). 

The D(4000) values do not appear to correlate neither with the absolute
magnitudes of the AGNs, which cover the range -23.9$<$M$_{V}$$<$-20.7, nor with the presence of broad
emission lines.  The histogram of the D(4000) index of the RDS AGNs is shown in Fig. 2a. This distribution
peaks at D(4000)$\sim$1.2, with most of the objects in the
range 1.0$\le$D(4000)$\le$1.4, plus some objects at larger values
(1.6$\le$D(4000)$\le$2.0). These two ranges of D(4000) correspond to galaxies with a sustained amount of star 
formation and galaxies with a dominant old stellar
population, respectively. The D(4000) values of the three group 
members (41C, 58B and 67B) are all consistent with those of elliptical galaxies.

As a comparison, the mean D(4000) 
values measured from the Canada-France Redshift Survey
(CFRS; Lilly et al. \cite{Li95}) in a redshift interval similar
to that of our objects (z$<$1) are 1.43 with a dispersion of 0.31 for 256
emission line galaxies and 1.92 with a dispersion of 0.45 for 85 "quiescent"
galaxies, defined as galaxies with no detected [OII] line (from Table 1
in Hammer et al. \cite{Ham97}). Similar average values (1.40 and 1.83 for emission
line galaxies and no-emission line galaxies) have recently been measured
from the large, more local (z$<$0.15) Stromlo-APM Survey (Tresse et al. \cite{Tre99}).

\begin{figure}[htb]
     \begin{minipage}{80mm}
     \psfig{figure=1729_f6.epsi,width=80mm,angle=-90,clip=}
     \end{minipage}
     \begin{minipage}{80mm}
        \psfig{figure=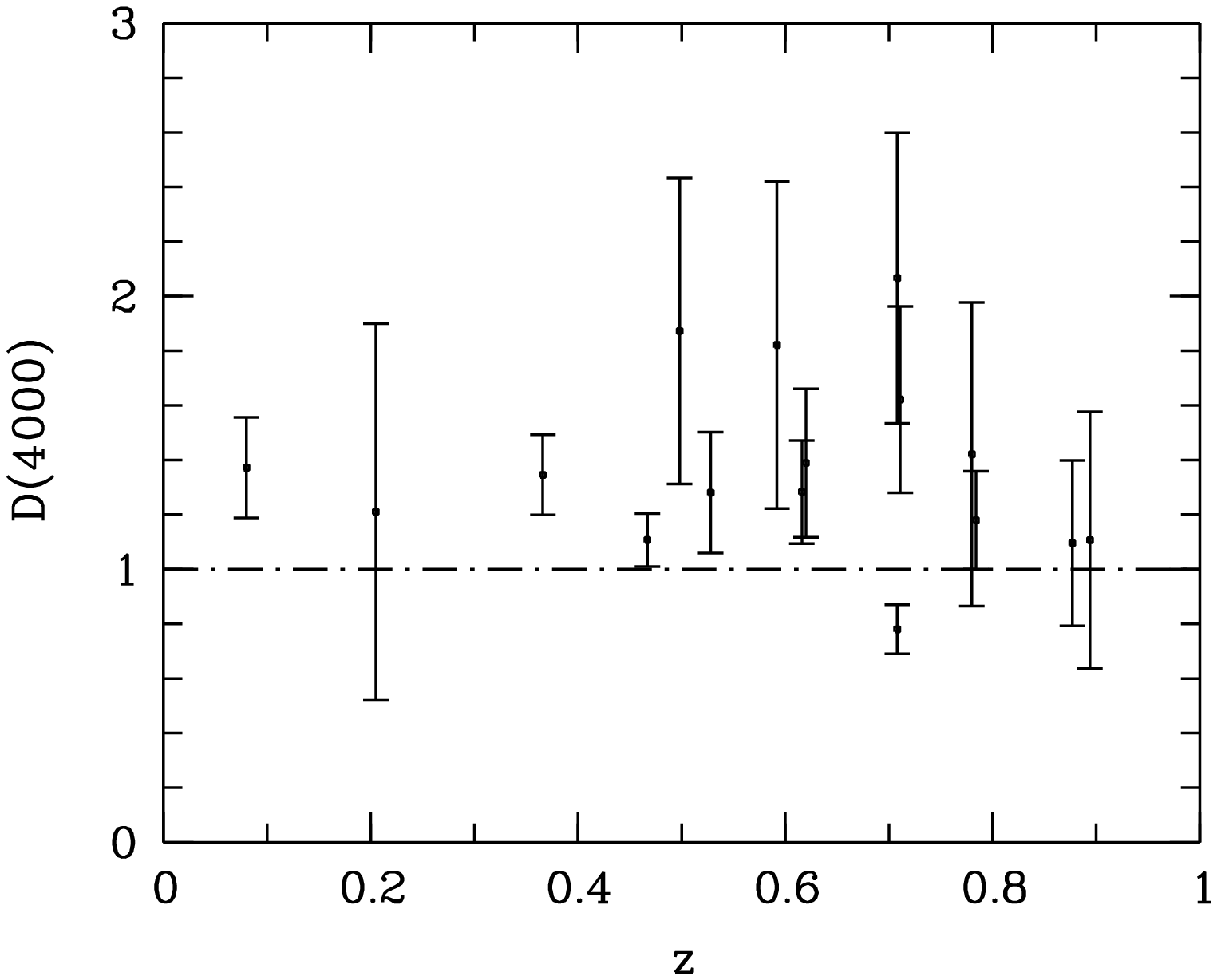,bbllx=46pt,bblly=85pt,bburx=400pt,bbury=525pt,width=80mm,angle=-90,clip=}
     \end{minipage}
     \caption[]{a) Histogram of the D(4000) index of the 16 RDS AGNs including 
                the values of the RDS galaxy 53A and of the group members 
                41C, 58B and 67B(s), which are marked with a different hatch. 
                b) Plot of the D(4000) index of the RDS AGNs (1$\sigma$ 
                errors) vs. redshift.
                About 63 \% of these AGNs show a significant galaxy component
                in addition to the AGN component (objects above the dash-dotted
                line).}
               \label{d4000}
\end{figure}
%

\section{New identifications}

The work described in Paper II was unable to identify 4 of the 50 sources:
14, 36, 84 and 116. We have obtained new optical and infrared 
observations of three of these objects, and have made one certain and two
probable identifications. In addition, the radio detection of source 116
suggests a tentative identification with a radio galaxy, which is probably an AGN.  
 
Table \ref{TAB2} gives the optical properties of the newly identified
X-ray sources in the same format as presented for the other sources 
in Paper II.  The first two columns
contain the name of the X-ray source and their flux in units of 10$^{-14}$
 erg cm$^{-2}$ s$^{-1}$ in the 0.5-2 keV band. The columns 3--6 give the name of the optical object identified with the X-ray source, its magnitude $R$, and its right ascension and declination at epoch 2000. The column 7 gives the distance of the optical object from the X-ray source in arcsec, which is based either on the 1112 ksec HRI pointing (H), or on the HRI raster scan (R) or on the 207 ksec PSPC exposure (P). The columns 8--10 contain the ratio f$_{x}$/f$_{v}$ as defined by Stocke et al.  (\cite{Sto91}), the morphological parameter (g$=$galaxy) and the redshift. The columns 11 and 12 present the optical absolute magnitude $M_{V}$ and the logarithm of the X-ray luminosity in the 0.5--2.0 keV energy band in units of erg s$^{-1}$. The optical classification 
and the ID class (see Sect. 3.) are given in the columns 13 and 14. 

\subsection{Optical identification of source 36}

The X--ray source 36 was identified by additional spectroscopic work in March 1998. The finding chart of 36 shows three faint optical objects within the PSPC error circle (see Fig. \ref{spec1}). The spectrum of object 36D shows two broad emission lines, identified as C III] and Mg~II at $z=1.524$. The object is therefore an AGN of the ID~class~a. 
The object 36A is too faint to obtain a reliable spectrum within 
an exposure time of 1800 sec, whereas the object 36E has a faint narrow 
emission line [O II] at $z=0.409$ indicating a normal NELG field object 
(marked with F in Fig. \ref{spec1}). Therefore 36D is likely to be the optical
counterpart of the X--ray source. The source south of 36D is a separate
X--ray source (not in this sample), which is optically identified with 36F, an AGN at $z=0.8$.

%
%
   \begin{table*}[t]
     \caption{Photometric and spectroscopic properties of the new identifications.}
      \label{TAB2}
      \begin{flushleft}
      \begin{tabular}{lccccc@{\hspace*{3mm}}cc@{\hspace*{2mm}}cccccc}
      \hline\noalign{\smallskip}
N$_{x}$ & S$_{x}$ & name & $R$ & $\alpha_{2000}$ & $\delta_{2000}$ & $\triangle$ pos & log f$_{x}$/f$_{v}$ & $G$ &  $z$  &  $M_{V}$  &  log L$_{x}$  &
class & ID class\\ \noalign{\smallskip}
(1) & (2) & (3) & (4) & (5) &  (6) &  (7) & (8) & (9) & (10) & (11) & (12) &(13) & (14) \\ \noalign{\smallskip}
      \hline\noalign{\smallskip}
      \noalign{\smallskip}
36 & 0.93 & 36D & 22.5 & 10 52 25.2 & 57 23 05.0 & 2\arcsec (R) & 0.71 & g & 1.524 & -22.1 & 44.14& AGN  & a\\
14 & 0.72 & 14Z & 25.0 & 10 52 42.4 & 57 31 58.4 & 2\arcsec (H)& 1.70 & g &-&-&-& AGN: & e\\
84 & 0.60 & 84Z &$>$25.5&10 52 17.1 & 57 20 16.8 & 2\arcsec (H)& $>$1.91 & g &-&-&-& AGN: & e\\
116& 0.57 &116A & 20.9 & 10 52 37.4 & 57 31 04.1 & 3\arcsec (P)& $-$0.57 & g & 0.708 & -21.9 & 43.25 & AGN  & e\\
      \noalign{\smallskip}
      \hline
      \end{tabular}
      \end{flushleft}
   \end{table*}

\subsection{Near--infrared identification of source 14 and 84}

We carried out deep broad-band K$^{\prime}$ (1.9244--2.292$\mu$m)
imaging of the two unidentified X-ray sources, 14 and 84, in 
December 1997. We used the Omega-Prime camera (Bizenberger et al. \cite{Biz98}) 
on the Calar Alto 3.5-m telescope. The camera uses a 1024$\times$1024 pixel 
HgCdTe HAWAII array with an image scale of 0.396 arcsec/pixel, and covers
a field-of-view of 6.7$\times$6.7 arcmin.  

Large numbers of short ($\sim$3 second) background-limited images were
taken at slightly dithered positions, with total accumulated integration
times of 45 minutes for X-ray source 14, and 40 minutes for source 84.
Standard techniques were used to reduce the data. For an image at any
given position, a series of images taken before and after at different 
positions were median combined to create a blank sky, which was then
subtracted from the source image. The result was then divided by a
tungsten-illuminated dome flat. Finally, all the reduced source images 
were registered and stacked to create the deep image. The integrated 
long-term seeing after combining all the images was $\sim$1 arcsec FWHM\@. 

%
\begin{figure}[ht]
     \begin{minipage}{42mm}
\centerline{\psfig{figure=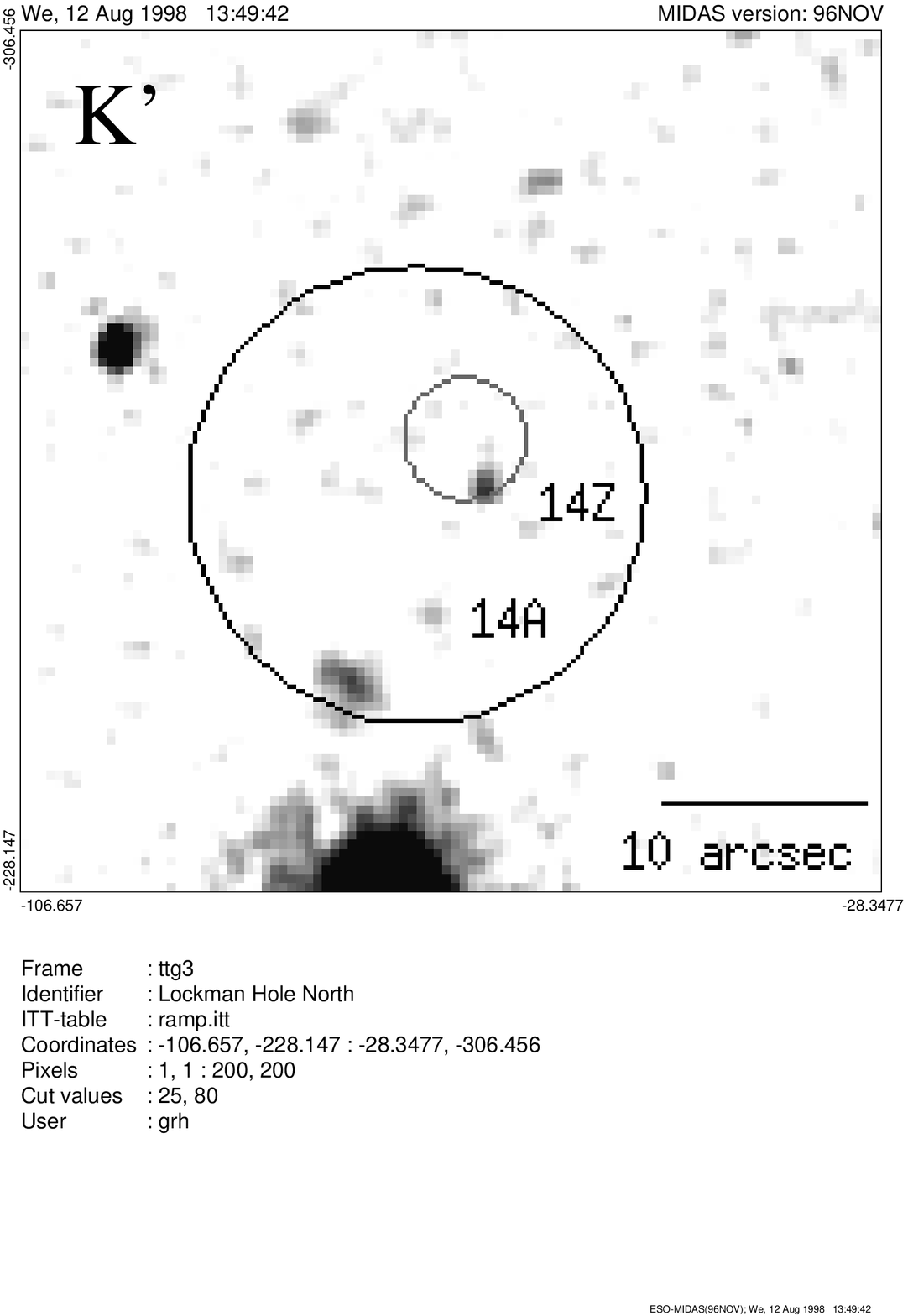,bbllx=11pt,bblly=220pt,bburx=447pt,bbury=654pt,width=42mm,clip=}}
     \end{minipage}
     \hfill
     \begin{minipage}{42mm}
\centerline{\psfig{figure=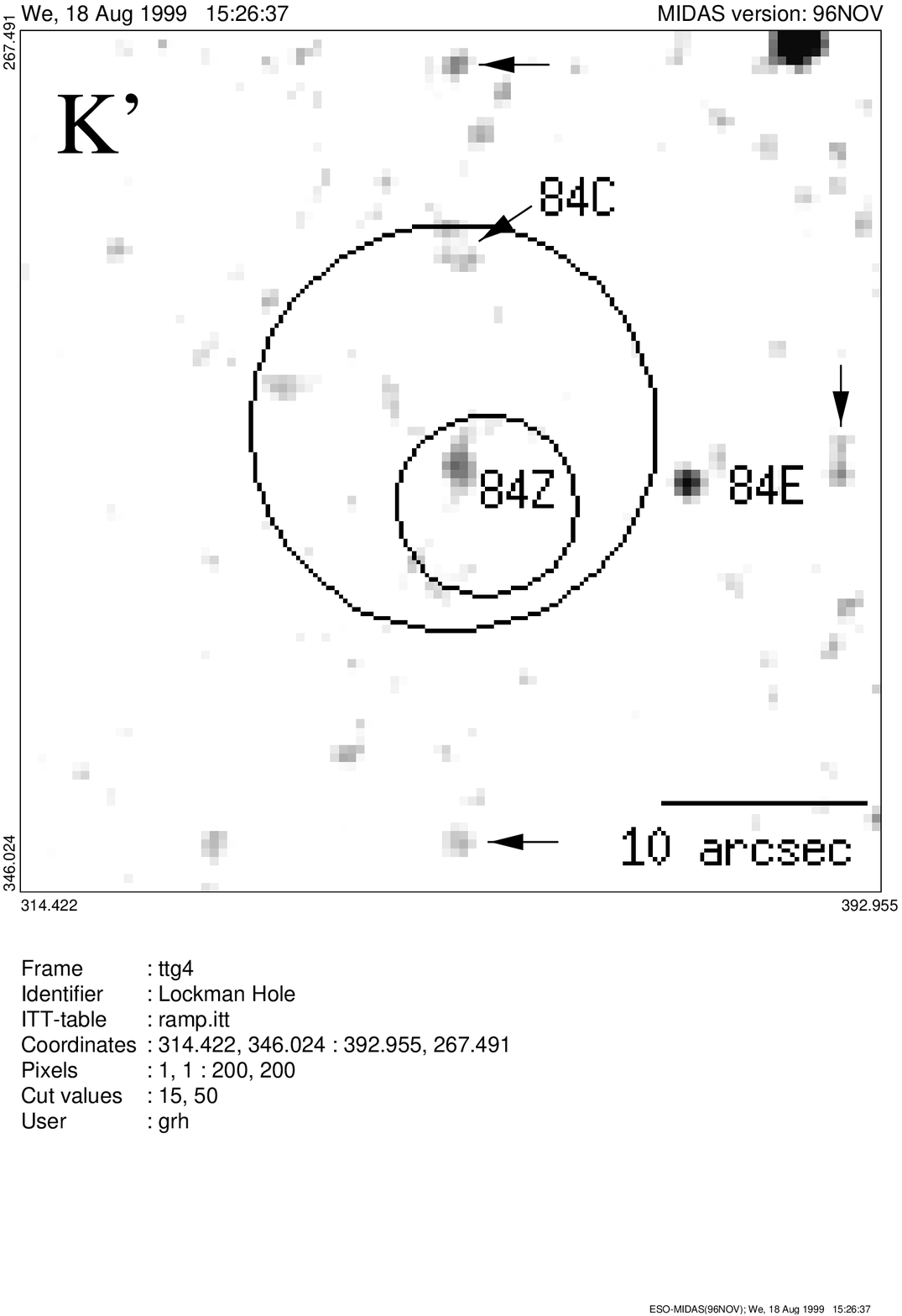,bbllx=15pt,bblly=220pt,bburx=447pt,bbury=654pt,width=42mm,clip=}}
     \end{minipage}
     \caption[]{$K$'-band images of the X-ray sources 14 and 84. North is to the top and east to the left. The PSPC and HRI error circles are indicated by the large
 and small error circles. Both X-ray sources have a $K$'-band counterpart within the HRI error circle whereas the $R$-band images show a very faint counterpart (14Z) or no optical object (84Z) at a limiting magnitude of $\sim$25.5 (see Fig. \ref{spec1}).}
     \label{Kband}
\end{figure}
%

%
The deep images were then smoothed with a Gaussian ($\sigma = 0.8$ pixels)
before sources were identified and their fluxes measured using the SExtractor package (Bertin \& Arnouts \cite{Ber96}). In the 45 minutes
exposure, point sources are well detected at the 5$\sigma$ level (peak pixel
brightness above the sky noise) at K$^{\prime}$=19.7 mag. 

Small subsections of the two $K^{\prime}$ images are presented in Fig.~\ref{Kband},
one each for sources 14 and 84. Faint $K^{\prime}$ sources are recognizeable
within the HRI error circle for both X-ray sources (14Z: $K^{\prime}=19.6$ mag, 84Z: $K^{\prime}=19.3$ mag).
New images taken with the Near Infrared Camera (NIRC; Matthews \& Soifer \cite{Matt94}) at the Keck telescope confirm the existence of this very red object (Schmidt, Soifer \& Thompson, priv. communication). 
 Both sources have a positional offset from the center of the HRI error circle of about 1\arcsec. Only the object 14Z was detected in the $R$-band image (cf. Fig. \ref{spec1}), whereas object 84Z shows no optical counterpart on the $R$-band image down to the limiting magitude of $R\sim25.5$ mag, which results in $R-K^{\prime}$ colours of $\sim5.4$ for 14Z and $\ge6.2$ for 84Z. 

The relatively low surface density of objects
with such red $R-K^{\prime}$ colours at K$^{\prime}$ $\sim19.5$ ($\sim0.5$ objects/sq.arcmin; 
see Fig 2. in Thompson et al. (\cite{Tho99}) strongly suggests that the
red sources near the HRI positions are the true counterparts of the X-ray
sources. The most probable identifications of such red objects are with either
high redshift (z $\ge$ 1) ellipticals or dust-reddened starbursts or AGNs.
In the first case, the X-ray emission would probably arise from a cluster; in the second case the dust-reddened AGN would have colors similar to those of a number of red quasars discovered in a radio selected sample (Webster et al. \cite{Web95}). 
Alternatively, they could be very high redshift AGNs (R-dropouts).
Since both X-ray sources have relatively hard spectra (Schmidt et al. \cite{Schm98};
see also Fig. 7 in Zamorani et al. \cite{Zam99}), indicative of a non-negligible
amount of X-ray absorption, we prefer the identification with a quasar reddened
by absorption for both sources. This identification is more secure for object
14Z, since no additional red object is seen close to it in the K$^{\prime}$ image (the
object 14A, inside the PSPC error box, but outside the HRI circle, is a NELG at
$z=0.546$, which is typical for the population of field galaxies (cf.
Sect. 6)), while it is more uncertain for object 84Z for which we can not
exclude the identification with a high redshift cluster. In fact, the K$^{\prime}$ image
shows a number of red objects (marked with arrows in Fig. 3), although with
less extreme $R-K^{\prime}$ colours than that of 84Z.
The objects 84C and 84E have been identified with NELG at redshifts of 0.381 and 0.525. A further NELG (84D) was found at $z=0.525$ (cf. Fig. \ref{spec1}). 

The large $f_{x}$/$f_{v}$ ratios of 14Z and 84Z (see Table \ref{TAB2}) are consistent with the ratios for AGNs, clusters of galaxies and BL Lac objects found by Stocke et al. \cite{Sto91} in the Einstein Medium Sensitivity Survey (EMSS). 

\subsection{Radio identification of source 116} 

Object 116A (z$=$0.708) near the center of the PSPC circle shows only absorption line features in its optical spectrum (cf. Fig. \ref{spec1}). Unfortunately, the wavelength range is not large enough to cover the regions of the H${\beta}$ and H$\alpha$ emission lines. The galaxies 116B and 116D near the edge of the PSPC circle are NELG at redshifts of 0.469 and 0.607. We have not yet obtained spectra of the galaxies 116C and 116E. 

The object 116A is close to the geometric center of two relatively bright radio sources which were detected at 20 cm and were considered to be the radio lobes of a double radio source by De Ruiter et al. (\cite{deRui97}; radio source \#99 in their Table 1). Recently, higher resolution VLA observations at 6 cm (Ciliegi, priv. communication) have detected a third radio source positionally coincident with the object 116A. The radio emission from this source appears to
be clearly connected with the two radio lobes. On this basis, we identify 116A with a double lobed radio--galaxy (AGN).

\section{Emission line properties}

In order to derive a reliable optical identification of the faint X--ray 
sources we have determined the emission line properties of the most prominent
AGN and galaxy emission lines. Due to the large redshift range of the RDS AGNs
from 0.080 to 2.832 with its mean redshift of $\left<z\right>=1.2$ (cf. Fig \ref{red}), any given emission line is present only in a small number of objects. Therefore we do not perform a correlation between the emission lines properties.
 
The emission lines have been measured by fitting 
with single or double Gaussian profiles applying the Levenberg-Marquardt 
algorithm (Press et al. \cite{Press92a}). First we fitted each
line with a Gaussian with four adjustable parameters; the total line flux,
the mean wavelength, the sigma of the Gaussian and the flux of a local linear
continuum. The Levenberg-Marquardt algorithm provides one sigma errors 
of the parameters. In some cases a single Gaussian gave a poor fit to the data, so an additional Gaussian was used (eg. for H$\beta$ of 37A or for H$\alpha$ of 59A).
%
%
   \begin{table*}[t]
      \caption{FWHM of UV emission lines in [km s$^{-1}$], corrected for instrumental resolution.}
      \label{TAB3}
      \begin{flushleft}
      \begin{tabular}{lccrrrrrr}
      \hline\noalign{\smallskip}
name &ID&$z$&Ly$_{\alpha}$~$\lambda$1216 & SI IV~$\lambda$1397 &C IV~$\lambda$1548 & C III]~$\lambda$1908 & Mg II~$\lambda$2798 & [Ne V]~$\lambda $3426 \\ 
\noalign{\smallskip}
(1) & (2) & (3) & (4)\hspace{0.5cm} & (5)\hspace{0.6cm} & (6)\hspace{0.5cm} & (7)\hspace{0.6cm} & (8)\hspace{0.6cm} & (9)\hspace{0.6cm} \\ \noalign{\smallskip}
\hline\noalign{\smallskip}
2A   &a&1.437 &\point &\point &\point & 870$\pm$40 & 6900$\pm$220 & 790$\pm$70 \\
6A   & a &1.204 & \point & \point & \point & 3560$\pm$20 & 1850$\pm$20$^{b}$  & 770$\pm$20 \\
     & & & & & & & 15920$\pm$180$^{v}$ &     \\
9A   &b&0.877  &\point &\point &\point &\point & 3380$\pm$100 &\point  \\
11A  &a&1.540  &\point &\point &\point & 1090$\pm$100 & 3490$\pm$100 &\point \\ 
12A$^{\ast}$ &d&0.990 &\point &\point &\point &\point &\point & 250$\pm$60 \\ \noalign{\smallskip}
16A  & c&0.586 &\point &\point &\point &\point &\point & $<$820$^{\ast\ast}$ \\
     & & & & & & & & 620$\pm$50\\
17A  &a&2.742 & 2240$\pm$10 & 4820$\pm$200 & 5600$\pm$40 & 1840$\pm$300$^{m}$ &\point  &\point \\
19B  &b&0.894 &\point &\point &\point &\point & 4240$\pm$60 &\point \\ 
23A$^{\ast}$ & b&1.009 &\point &\point &\point &\point & 3630$\pm$50 &\point \\ 
25A  &a&1.816 &\point &\point &\point & 4840$\pm$410 & 5790$\pm$250 &\point \\ \noalign{\smallskip}
27A  &a &1.720 &\point &\point &\point & 11760$\pm$820$^{v}$ & 3810$\pm$260 &\point \\
29A  & b&0.784 & \point &\point &\point & \point & 8120$\pm$20$^{v}$ & 1050$\pm$60 \\
30A  &a&1.527  &\point &\point & 3940$\pm$70 & 5260$\pm$80 & 3020$\pm$40 & \point\\ 
31A  &a &1.956 &\point & 5630$\pm$70 & 2880$\pm$10$^{b}$ & 4670$\pm$30 &\point &\point \\ 
   & & & &  & 11060$\pm$70$^{v}$& & & \\ 
32A  &a &1.113  & \point &\point &\point  & 1020$\pm$20$^{n}$ & 2560$\pm$10$^{b}$  & 900$\pm$20 \\ 
     & & & & &  & 6740$\pm$40$^{b}$ & 15020$\pm$70$^{v}$ &     \\ \noalign{\smallskip}
35A  &a&1.439  &\point &\point &\point & 7460$\pm$30 & 8640$\pm$20$^{v}$ &\point \\
36D  &a&1.524  &\point &\point &\point &3000$\pm$180& 4730$\pm$160 & \point\\
37A  &b&0.467  &\point &\point &\point &\point & 1810$\pm$70 &\point  \\
37G  &a&2.832  & $<$480 &5170$\pm$120 & 4380$\pm$30 & 5800$\pm$240 &\point &\point \\ 
    &   & & 2800$\pm$40$^{b}$ &      &      &      & &  \\  
38A$^{\ast}$  &a&1.145 &\point &\point &\point & 4520$\pm$2350$^{m}$ & 7320$\pm$450 &\point \\  \noalign{\smallskip}
42Y  &a&1.144 &\point  &\point &\point & 7320$\pm$100 & 5260$\pm$40 & \point\\
43A$^{\ast}$  &a&1.750 &\point &\point &\point & 210$\pm$60$^{m}$ & $\sim4800$$^{e}$ &\point \\
46A$^{\ast}$  &a&1.640 &\point &\point & 9120$\pm$260$^{v}$ & 6000$\pm$160 & 1120$\pm$110 &\point \\
47A  &a&1.058 &\point &\point &\point & 5400$\pm$180 & 1910$\pm$100 &\point \\ 
51D  &d&0.528 &\point &\point &\point &\point &\point & 480$\pm$60 \\ \noalign{\smallskip}
51L  &d&0.620 & & & & & & 2780$\pm$180$^{\rm \ast\ast\ast}$ \\
52A  &a&2.144 &\point &\point & 6220$\pm$480 & 6040$\pm$840 &\point &\point \\
54A  &a&2.416 & 1160$\pm$20 & 3510$\pm$170 & 1850$\pm$30$^{b}$ & 5050$\pm$60 &\point &\point \\
     & & &              &              & 8460$\pm$100$^{v}$&             & & \\
55C  &a&1.643 &\point &\point & 5260$\pm$20 & 7880$\pm$60 & 4020$\pm$20 & \point\\ 
56D  &b&0.366 &\point &\point &\point &\point & 2460$\pm$40 & $<$610 \\ \noalign{\smallskip}
60B  &a&1.875 &\point &\point & 1760$\pm$30 & 2500$\pm$50 & 3850$\pm$90 &\point \\
61B  &b&0.592 &\point &\point &\point &\point & 6620$\pm$440 & \point \\
73C  &a&1.561 &\point &\point & 4340$\pm$120 & 4610$\pm$130 & 4520$\pm$40 &\point \\
77A  &a&1.676 &\point &\point & 8450$\pm$230$^{v}$ &\point & 6730$\pm$130 & \point\\ 
117Q &d&0.780 &\point &\point &\point &\point &\point & 560$\pm$70 \\
       \noalign{\smallskip}
      \hline
      \end{tabular}
      \end{flushleft}
          \begin{list}{}{}
          \item[$^{\rm \ast)}$] More than one spectrum available, $^{\rm \ast\ast)}$ Emission line [NeV]$\lambda$3346, $^{\rm \ast\ast\ast)}$ Probably blended with the 5570 \AA~sky line.
          \item[$^{\rm m)}$] Marginal line detection ($2-3\sigma$), $^{\rm e)}$ estimated from the heavily smoothed spectrum.
          \item[$^{\rm n)}$] Narrow, $^{\rm b)}$ broad and $^{\rm v)}$ very broad component (FWHM$>$8000 km s$^{-1}$) of the emission line.
          \end{list}
      \end{table*}
%
   \begin{table*}[t]
      \caption{FWHM of optical emission lines in [km s$^{-1}$], corrected for
      instrumental resolution.}
      \label{TAB4}
      \begin{flushleft}
      \begin{tabular}{lccrrrrrrr}
      \hline\noalign{\smallskip}
name &ID & $z$ & [O II]~$\lambda$3726 & [Ne III]~$\lambda$3868 & H$_{\gamma}$~$\lambda$4340 & H$_{\beta}$~$\lambda$4861 & [O III]~$\lambda$4959 & [O III]~$\lambda$5007 & H$_{\alpha}$~$\lambda$6563 \\ \noalign{\smallskip}
(1) & (2) & (3) & (4)\hspace{0.6cm} & (5)\hspace{0.6cm} & (6)\hspace{0.4cm} & (7)\hspace{0.4cm} & (8)\hspace{0.7cm} & (9)\hspace{0.7cm} & (10)\hspace{0.4cm} \\ \noalign{\smallskip}
\hline\noalign{\smallskip}
9A    &b&0.877 & $<$830 &\point &\point &\point &\point & \point&\point \\
12A$^{\ast}$   &d&0.990 & 280$\pm$20 &\point &\point &\point &\point &\point &\point \\
16A   &c&0.586 & $<$550 & $<$670 & $<$660 & 2980$\pm$20 & $<$240 & 320$\pm$10 &\point \\
19B   &b&0.894 & 510$\pm$50 &\point &\point &\point &\point &\point &\point \\
23A$^{\ast}$   &b&1.009 & 390$\pm$40 &\point &\point &\point &\point & \point&\point \\ \noalign{\smallskip}
26A$^{\ast}$   &d &0.616 & 330$\pm$10 & 1220$\pm$40 & $<$390 & 250$\pm$10 & 280$\pm$30 & 280$\pm$10 &\point \\
28B   &c &0.205 &\point &\point &\point & 2510$\pm$330  & $<$1050 & $<$1180 & 3360$\pm$30 \\
29A   &b &0.784 & 300$\pm$20 & 630$\pm$30 & 290$\pm$30 &\point &\point &\point &\point \\
32A   &a&1.113 & 200$\pm$10 &\point &\point &\point &\point &\point &\point \\
37A   &b &0.467 & $<$450 &\point & 890$\pm$20 & $<$350$^{n}$ & $<$360 & $<$360 &\point \\ 
      &  & & & &     & 2070$\pm$20$^{b}$ & & & \\ \noalign{\smallskip}
45Z$^{\ast}$   &d &0.711 & 540$\pm$10 & 1270$\pm$60 &\point &\point &\point &\point &\point \\
48B$^{\ast}$   &e&0.498 & 930$\pm$40  &\point &\point &\point &\point &\point &\point \\
48Z   & &0.502  & 570$\pm$30 &\point &\point & 250$\pm$70 &\point & 580$\pm$60 & \point\\
51D   &d&0.528  & 450$\pm$20 &\point &\point &\point & 370$\pm$60 & 300$\pm$50 &\point \\ 
51L   &d &0.620 & $<$480 & $<$340  & $<$280  & $<$320  & 220$\pm$20 & $<$340 &\point \\ \noalign{\smallskip}
53A   &e &0.245 &\point &\point &\point &\point &\point & $<$390 & $<$310 \\
56D   &b &0.366 & $<$510 & 340$\pm$60 & & 7680$\pm$70 & 190$\pm$40 & 220$\pm$10 &\point \\
58B  &e&0.629 & $<$470 &\point &\point &\point &\point &\point &\point \\
59A   &c &0.080 & $<$580 & $<$490 & & $<$410 & 840$\pm$40 & $<$550 & $<$300$^{n}$ \\ 
      &  & & &   & &   &     &        & 8320$\pm$330$^{v}$ \\
61B   &b &0.592 & 310$\pm$50 &\point &\point &\point & \point& 430$\pm$30 &\point \\ \noalign{\smallskip}
67A   &e &0.546 & 210$\pm$40 &\point &\point & $<$320 &\point & $<$390 &\point \\
117Q  &d &0.780 & 550$\pm$10 & 630$\pm$40 & \point&\point &\point &\point &\point \\
      \noalign{\smallskip}
      \hline
       \end{tabular}
      \end{flushleft}
      \end{table*}
%

From the best set of parameters we have calculated the redshift, the rest frame EW$_{rest}$ in \AA~and the FWHM in km s$^{-1}$. The redshift of each object was derived as the mean value from the strongest features (eg. Mg II $\lambda$ 2798, [O II] $\lambda$ 3727). Its mean error is around $\sim$0.001. The FWHM has been corrected for intrumental resolution, using 
10 \AA~and 15 \AA~for the mean instrumental resolution of the 
Keck long-slit and multi-slit spectra and 25 \AA~for the Palomar spectra.
The mean instrumental resolution of the spectra was obtained by fitting 
a Gaussian to the emission lines of the wavelength calibration spectra.
If an emission line was unresolved (F$_{obs} \simeq$ W), we set W$=$0 \AA~to derive an upper limit of the FWHM [km~s$^{-1}$]. The largest possible error of FWHM was calculated assuming that the error of the mean instrumental resolution $\sigma_{\mathrm{W}}$ is zero. We have determined for some objects the FWHM and its error from several observations (eg. 14A, 43A). These values are generally in good agreement within errors. Only for the object 38A there is a significant difference in the individual values of the CIII] emission line. Therefore we used the mean value of both measurements, which has a relatively large error.
The FWHM of all UV and optical emission lines is given in Table \ref{TAB3} and in Table \ref{TAB4}, respectively. The first two columns contain the name of the optical counterpart and its ID class (for description see Sect. 4). The third column lists the mean redshift of the object. The columns 4--10 give the FWHM [km s$^{-1}$] of the prominent UV or optical emission lines, if found in the spectra. The values are rounded to the nearest integer that can be divided by 10. 
In a few cases the two Gaussian fits for the UV lines C IV, CIII] and Mg II required both a broad and a very broad component (FWHM$>$8000 km s$^{-1}$), which is indicated with ''v'' in Table \ref{TAB3}.
In Sect. 5.1 we discuss the FWHM of AGNs in the RDS in comparison with the values found in other X--ray surveys.

Table \ref{TAB5} and Table \ref{TAB6} present the rest frame EW of the 
UV and optical emission lines of the RDS sources. The format is the same as for the FWHM (Table \ref{TAB3} and Table \ref{TAB4}) except for the third column, which contains the optical classification of the counterpart (AGN, group or galaxy) instead of the ID class.
The EW$_{rest}$ values derived from the Keck spectra have on average much smaller errors than those derived from the Palomar spectra (9A, 25A, 27A, 28B and 52A). This is a consequence of the better spectral resolution of the Keck spectra. In Sect. 5.2 we discuss the rest frame EW of RDS AGNs.

\subsection{FWHM of the RDS AGNs}

First we have divided the emission lines into narrow emission lines ($<$1500 km s$^{-1}$) and broad emission lines ($>$1500 km s$^{-1}$). In addition, a few spectra required very broad line components ($>$8000 km s$^{-1}$) of the
semi--permitted line C III] and of the permitted lines C IV and Mg II.
The unresolved narrow lines and marginally detected lines (cf. Table \ref{TAB3} and \ref{TAB4}) were neglected in the following analysis. Since there are only few observations of  Ly$\alpha$ and H$\alpha$ we do not present a statistical analysis for these lines.
If there were at least three values for narrow or broad components available, we have calculated a mean and an error on the mean (the standard deviation $\sigma$ divided by the square root of the number n). We compare the results (FWHM) of the RDS AGNs with the properties of the UV/optically selected, low--redshift quasar sample ($\left<z\right>=0.54$) from Green (\cite{Gre96}; hereafter GR) and with the FWHM of a QSO sample at intermediate redshift (0.9$\leq z \leq$2.2) from Brotherton et al. (\cite{Bro94}; hereafter BR). 
In addition we used for comparison the emission line properties derived from two X--ray surveys: the RIXOS AGN X--ray sample with $\left<z\right>=0.82$ (Puchnariwicz et al. \cite{Puch97}) and the AGN data of the Cambridge--Cambridge ROSAT Serendipity Survey (CRSS) of Boyle et al. (\cite{Boy97}), which has a mean redshift of $\left<z\right>=0.88$.
We have subdivided the emission lines into forbidden lines ($[$NeV$]$, $[$O~II$]$, $[$Ne~III$]$ and $[$O~III$]$), Balmer lines (H${\gamma}$, H${\beta}$ and H${\alpha}$) and typical quasar lines (
Ly${\alpha}$, SI~IV, C~IV, CIII$]$ and Mg~II). \\
\hspace*{5mm} Because many RDS AGNs show a strong continuum from the host galaxy (see Sect. 3.1), we have searched for differences in the emission line properties of the forbidden lines of the RDS AGNs and of our field NELG, which are physically not associated with the RDS X--ray sources as described in Sect. 6. Therefore we compare the FWHM and the rest frame EW of the RDS AGNs in addition to the AGN comparison samples with those values derived from our field NELG sample.  

Quasar lines: Table \ref{TAB8} contains the mean FWHM, the standard deviation, the error on the mean (column 2--4) and the number of RDS quasar lines (column 5). The values in brackets give the FWHM including the very broad emission lines (FWHM$>$8000 km s$^{-1}$). In addition, we list the mean FWHM with its error and the number of objects in parentheses of the RIXOS, the CCRS, the GR and the BR sample in columns 6--9. The mean SI~IV FWHM of the four RDS AGNs is consistent with the GR data. The mean RDS C~IV FWHM agrees quite well with that of the CCRS sample and with the mean FWHM, which we have derived from the BR data. Green (\cite{Gre96}) has obtained a significantly larger FWHM for his low-redshift quasar sample.
The mean C~III$]$ FWHM (broad component) of the RDS AGNs is between the value of the CCRS sample and the values of several AGN samples (RIXOS, GR, BR).  
Furthermore, three C~III$]$ narrow emission lines were detected, which have a mean FWHM 990$\pm$65 ($\sigma$$=$110) km s$^{-1}$. The mean Mg~II FWHM, including very broad components, of RDS AGNs is similar to, although slightly higher than, those from the CCRS sample, the RIXOS survey and the BR sample. 
The object 54A shows a narrow  Ly${\alpha}~\lambda1216$ emission line of 1160 km s$^{-1}$. The FWHM of the broad Ly${\alpha}~\lambda1216$ emission lines with 2240 km s$^{-1}$ (17A) and 2800 km s$^{-1}$ (37G) are at least a factor of 2 lower than found for the GR sample ($7690\pm250$~km s$ ^{-1}$ with $\sigma$$=$2200 km s$ ^{-1}$). 

%
   \begin{table*}[t]
      \caption{Distribution of the FWHM [km s$^{-1}$] of the RDS quasar lines.}
      \label{TAB8}
      \begin{flushleft}
      \begin{tabular}{lcccccccc}
      \hline\noalign{\smallskip}
\hspace{0.5cm}lines    &  \multicolumn{4}{c}{RDS} & RIXOS  & CCRS  & GR  & BR \\
   & mean &  $\sigma$ & error & n & mean &  mean & mean  &mean \\
\hspace{0.5cm}(1) & (2) & (3) & (4) & (5) & (6) & (7) & (8) & (9) \\ \noalign{\smallskip}
\hline \noalign{\smallskip}
SI~IV~$\lambda1397$  & 4780        & 910&450& 4&  - & - &4050$\pm$550 (70)& -\\
C~IV~$\lambda1548$   & 4030~(5640) &1590&530& 9&  -&5100$\pm$640 (14) &7820$\pm$360 (63) &4800$\pm$270 (43) \\
C~III$]~\lambda1908$ & 5380~(5660) &1550&390&16&6300$\pm$400 (42)&4800$\pm$340 (28)&6340$\pm$520 (52) &5980$\pm$260 (85)\\
Mg~II~$\lambda2798$  & 4190~(5420) &1710&370&21&4600$\pm$200 (132)&4700$\pm$370 (50) & -&4440$\pm$200 (67)\\
      \noalign{\smallskip}
      \hline
      \end{tabular}
      \end{flushleft}
      \end{table*}

   \begin{table*}[t]
      \caption{FWHM [km s$^{-1}$] distribution of the RDS AGNs and of the field NELG forbidden lines.}
      \label{TAB7}
      \begin{flushleft}
      \begin{tabular}{lccccc@{\hspace*{6mm}}|@{\hspace*{6mm}}ccccc}
      \hline\noalign{\smallskip}
\hspace{0.5cm}lines     &\multicolumn{4}{c}{RDS AGNs} & frac. &  \multicolumn{4}{c}{Field NELG}    & frac. \\
               & mean &  $\sigma$ & error & n  & & mean &  $\sigma$ & error & n &  \\
\hspace{0.5cm}(1) & (2) & (3) & (4) & (5) & (6) &(7) & (8) & (9) & (10)& (11)\\ \noalign{\smallskip}
\hline \noalign{\smallskip}
$[$Ne~V$]~\lambda3426$   &680 & 250 & 90 & 8  & 89\% &-&-&-&-&- \\
$[$O~II$]~\lambda3727$   & 440 & 200 & 60 & 11 & 65\% &290&120&30&18&30\% \\
$[$Ne~III$]~\lambda3868$ & 820 & 400 & 180 & 5 & 62\% & 770& 100&60&3&38\% \\
$[$O~III$]~\lambda4959$  & 380 & 270 & 120 & 5 & 62\% & 600&320&130&6&29\% \\
$[$O~III$]~\lambda5007$  & 310 &  80 &  30 & 5 & 56\%&290&100&40&7&24\%\\
      \noalign{\smallskip}
      \hline
      \end{tabular}
      \end{flushleft}
      \end{table*}
%

Balmer lines: The AGNs 28B and 59A exhibit broad H${\alpha}$ emission lines. The FWHM of 59A (8320 km s$^{-1}$) is larger than detected for any broad H${\alpha}$ component of the RIXOS sample, with its mean FWHM of 3100$\pm$400 ($\sigma$$=$1800) km s$^{-1}$. The mean FWHM of the broad H${\beta}~\lambda4861$ line 3810$\pm$1300 ($\sigma$$=$2610) km~s$^{-1}$ agrees with the values from other X--ray selected AGN samples (CCRS: 3400$\pm$160 ($\sigma$$=$2400) km~s$^{-1}$, RIXOS: 3900$\pm$400 ($\sigma$$=$2000) km~s$^{-1}$ and ST: 3640$\pm$230 ($\sigma$$=$1750) km~s$^{-1}$). Boroson \& Green (\cite{Bo92}; hereafter BG) have obtained a value of 3800$\pm$200 ($\sigma$$=$1860) km~s$^{-1}$ for a bright optically selected QSO sample at low redshift $\left<z\right>=0.17$. Only one of five narrow H${\beta}$ lines was resolved (cf. Table \ref{TAB4}). Two of five narrow H${\gamma}$ emission lines have been resolved, but none of them show a broad emission line component. 

Forbidden lines: Table \ref{TAB7} presents the FWHM of RDS AGNs and the values of field NELG. The first column is the name of the line. The columns 2--5 contain the mean FWHM, the standard deviation, the error on the mean and the number of emission lines for RDS AGNs. All the numbers in these columns refer to the resolved emission lines. Column 6 is the fraction of resolved emission lines compared with the total number of detected lines. The columns 7--11 show the same values for our sample of field NELG.

The fraction of resolved NELG emission lines ($\sim$30 \%) is smaller than the fraction ($\sim$60 \%) found for RDS AGNs, which is probably due to the higher S/N  ratio of the RDS spectra. The optical spectra of field NELG show no $[$Ne V$]$ emission lines as found for several RDS AGNs. The mean RDS FWHM and the mean field NELG FWHM of the forbidden lines are in good agreement within errors. The mean FWHM of the [Ne III] lines seems to be slightly larger than those of the oxygen lines. But this is probably due to a blend with other lines in this region of a noisy continuum.\\

To summarize, in general the RDS AGN FWHM of the broad emission lines are consistent with the values found for other X--ray selected AGN samples as well as with the data obtained from optical/UV selected AGN samples.

\subsection{Rest frame EW of the RDS AGNs}

In addition to the already mentioned AGN samples, we use the range of the mean EW derived for several emission line surveys (Osmer \cite{Os80}, Vaucher \& Weedman \cite{Vau80}) and optical/radio quasar surveys (Neugebauer et al. \cite{Neu79}, Richstone \& Schmidt \cite{Ri80}, Oke \& Korycanski \cite{Oke82}, Wampler et al. \cite{Wa84}, Tytler et al. \cite{Ty87}) given in Table 6 of Schmidt et al. (\cite{Schm86}; hereafter SSG) and the sample of high-redshift, bright QSO $\left<z\right>$$=$1.6 from Steidel \& Sargent (\cite{Stei91}; hereafter SS) for a comparison with the RDS AGN EW. Further, we have calculated the mean EW (neglecting uncertain values) from the X--ray selected AGN sample ($\left<z\right>$$=$0.19) of Stephens (\cite{Step89}; hereafter ST). 
 
Quasar lines: The mean EW of the RDS quasar lines are shown in Table \ref{TAB10}, where columns 2--5 give the mean EW, the standard deviation, the error on the mean and the number of objects. The values in brackets in column 2 include the very broad line components. Columns 6--9 contain the mean EW, its error and the number of objects of the comparison AGN samples. The column 10 gives the range of mean EW taken from Table 6 in SSG. The mean SI~IV EW of the RDS AGNs is about twice the value of the GR sample. The mean C~IV, C~III] and Mg~II EW of the RDS AGNs have relatively large errors due to the spread within a small number of emission lines, but they are in general good agreement, although on the high side, with the values of the comparison samples. Further they are in the range of mean EW values derived from several quasar surveys (SSG).  

%
   \begin{table*}
      \caption{Rest frame EW of UV emission lines in [\AA].}
      \label{TAB5}
      \begin{flushleft}
      \begin{tabular}{lccrrrrrr}
      \hline\noalign{\smallskip}
name & $z$ &class & Ly$_{\alpha}$~$\lambda$1216 & SI IV]~$\lambda$1397 &C IV~$\lambda$1548 & C III]~$\lambda$1908 & Mg II~$\lambda$2798 & [Ne V]~$\lambda$3426 \\ \noalign{\smallskip}
(1) & (2) & (3) & (4)\hspace{0.5cm} & (5)\hspace{0.6cm} & (6)\hspace{0.5cm} & (7)\hspace{0.6cm} & (8)\hspace{0.6cm} & (9)\hspace{0.6cm} \\ \noalign{\smallskip}
\hline\noalign{\smallskip}
2A   &1.437 &AGN &\point &\point &\point & 3.0$\pm$0.2$^{n}$ & 15.0$\pm$1.2$^{b}$& 1.5$\pm$0.3 \\
6A  &1.204&AGN &\point  &\point  &\point  & 17.0$\pm$0.3$^{b}$ & 10.6$\pm$0.3$^{b}$ & 1.8$\pm$0.1\\
    & &     &  & & &                & 56.5$\pm$1.9$^{v}$ &    \\
9A  &0.877&AGN &\point &\point &\point &\point & 57.7$\pm$4.7$^{b}$ &\point  \\
11A  &1.540 &AGN &\point &\point &\point & 9.5$\pm$1.9$^{n}$ & 19.2$\pm$1.5$^{b}$ &\point  \\ 
12A$^{\ast}$  &0.990&AGN &\point &\point &\point &\point &\point & 7.7$\pm$1.1 \\ \noalign{\smallskip}
16A &0.586 &AGN &\point &\point &\point &\point &\point & 0.8$\pm$0.2$^{\ast\ast}$ \\
    &  & & & & & &  & 3.1$\pm$0.2  \\
17A &2.742 &AGN & 55.8$\pm$1.0$^{b}$ & 8.8$\pm$0.9$^{b}$ & 49.7$\pm$1.3$^{b}$ & 5.0$\pm$2.0$^{m}$ &\point &\point \\
19B  &0.894&AGN &\point &\point &\point &\point & 39.8$\pm$1.5$^{b}$ &\point \\ 
23A$^{\ast}$  &1.009 &AGN &\point &\point &\point &\point & 46.6$\pm$1.9$^{b}$ &\point  \\
25A  &1.816 &AGN &\point &\point &\point & 27.2$\pm$6.0$^{b}$ & 53.6$\pm$7.0$^{b}$ &\point \\ \noalign{\smallskip}
27A  &1.720 &AGN &\point &\point &\point & 53.4$\pm$10.6$^{v}$ & 23.0$\pm$3.9$^{b}$ &\point \\
29A &0.784&AGN  &\point &\point &\point &\point & 144.0$\pm$1.3$^{v}$ & 5.4$\pm$0.6  \\
30A &1.527&AGN &\point &\point & 20.8$\pm$1.1$^{b}$ & 17.7$\pm$0.7$^{b}$ & 19.4$\pm$0.7$^{b}$ & \point \\ 
31A &1.956&AGN &\point & 17.9$\pm$0.6$^{b}$ & 46.1$\pm$0.9$^{b}$  & 17.3$\pm$0.3$^{b}$ &\point &\point \\
    &  & &             &   & 60.2$\pm$1.5$^{v}$  & &  \\
32A &1.113 & AGN &\point &\point &\point & 1.9$\pm$0.1$^{n}$ & 9.0$\pm$0.1$^{b}$ & 0.7$\pm$0.1 \\ 
    & &  & & & & 17.0$\pm$0.2$^{b}$ & 44.4$\pm$0.5$^{v}$ &  \\ \noalign{\smallskip}
35A  &1.439&AGN  &\point &\point &\point & 16.7$\pm$0.2$^{b}$ & 36.6$\pm$0.2$^{v}$ &\point \\ 
36D &1.524 &AGN &\point &\point &\point & 254.1$\pm$68.1$^{b}$& 262.1$\pm$31.5$^{b}$ &\point  \\
37A &0.467 &AGN &\point &\point &\point &\point & 10.8$\pm$0.9$^{n}$ & \point \\
37G &2.832 &AGN& 5.5$\pm$0.4$^{n}$ & 8.5$\pm$0.5$^{b}$ & 42.7$\pm$0.9$^{b}$ & 17.3$\pm$1.9$^{b}$&\point &\point \\
      & &    & 36.2$\pm$1.7$^{b}$                &                &                &   \\
38A$^{\ast}$ &1.145&AGN &\point &\point &\point & 15.4$\pm$6.1$^{m}$ & 39.9$\pm$2.3$^{b}$ &\point \\ \noalign{\smallskip}
42Y  &1.144 &AGN &\point &\point &\point & 18.3$\pm$0.6$^{b}$ & 35.5$\pm$0.6$^{b}$ & \point\\
43A$^{\ast}$ &1.750 &AGN &\point &\point &\point & 33.0$\pm$17.0$^{m}$ &\point &\point \\
46A$^{\ast}$ &1.640 &AGN &\point &\point & 117.3$\pm$34.0$^{v}$ & 43.4$\pm$3.6$^{b}$ & 165.0$\pm$40.0$^{n}$ &\point \\
47A &1.058 &AGN &\point &\point &\point & 82.5$\pm$10.7$^{b}$ & 13.9$\pm$1.8$^{b}$ &\point \\
51D  &0.528&AGN &\point &\point &\point &\point &\point & 2.7$\pm$0.3 \\ \noalign{\smallskip}
51L &0.620 &AGN &\point &\point &\point &\point &\point & 9.3$\pm$1.4 \\
52A  &2.144&AGN &\point &\point & 33.5$\pm$7.5$^{b}$ & 24.6$\pm$9.0$^{b}$ &\point &\point \\
54A  &2.416&AGN & 55.0$\pm$2.9$^{n}$ & 11.9$\pm$1.5$^{n}$ & 24.6$\pm$1.3$^{b}$ & 37.9$\pm$1.6$^{b}$&\point &\point \\
     & & &               &                & 76.1$\pm$4.3$^{v}$ &               & & \\
55C  &1.643 &AGN &\point &\point & 28.3$\pm$0.4$^{b}$ & 28.8$\pm$0.6$^{b}$ & 30.2$\pm$0.4$^{b}$ & \point\\ \noalign{\smallskip}
56D &0.366 &AGN  &\point &\point &\point &\point & 16.0$\pm$0.6$^{b}$ & 1.2$\pm$0.2  \\
60B &1.875 &AGN &\point &\point & 12.7$\pm$0.6$^{b}$ & 9.8$\pm$0.5$^{b}$ & 24.4$\pm$1.6$^{b}$ &\point \\
61B &0.592 &AGN &\point &\point &\point &\point & 115.3$\pm$29.7$^{b}$ & \point\\
73C  &1.561 &AGN &\point &\point & 77.9$\pm$10.1$^{b}$ & 29.3$\pm$2.4$^{b}$ & 47.5$\pm$1.2$^{b}$ &\point \\
77A  &1.676 &AGN &\point &\point & 115.2$\pm$15.4$^{v}$ &\point & 83.5$\pm$6.1$^{b}$ &\point \\ \noalign{\smallskip}
117Q &0.780&AGN &\point &\point &\point &\point &\point & 12.8$\pm$2.6 \\
      \noalign{\smallskip}
      \hline
      \end{tabular}
      \end{flushleft}
       \begin{list}{}{}
          \item[$^{\rm m)}$]Marginal line detection ($2-3\sigma$).
       \end{list}
      \end{table*}
%
%
%
   \begin{table*}
      \caption{Rest frame EW of optical emission lines in [\AA].}
      \label{TAB6}
      \begin{flushleft}
      \begin{tabular}{lcc@{\hspace*{2mm}}rrrrrrr}
      \hline\noalign{\smallskip}
name & $z$ &class& [O II]~$\lambda$3726 & [Ne III]~$\lambda$3868 & H$_{\gamma}$~$\lambda$4340 & H$_{\beta}$~$\lambda$4861 & [O III]~$\lambda$4959 & [O III]~$\lambda$5007 & H$_{\alpha}~\lambda$6563 \\ \noalign{\smallskip}
(1) & (2) & (3) & (4)\hspace{0.6cm} & (5)\hspace{0.6cm} & (6)\hspace{0.4cm} & (7)\hspace{0.4cm} & (8)\hspace{0.7cm} & (9)\hspace{0.7cm} & (10)\hspace{0.4cm} \\ \noalign{\smallskip}
\hline \noalign{\smallskip}
9A &0.877&AGN& 10.2$\pm$1.5 &\point &\point &\point &\point &\point &\point \\
12A &0.990&AGN&15.9 $\pm$1.0 &\point &\point &\point &\point &\point &\point \\
16A &0.586&AGN & 1.7$\pm$0.2 & 2.8$\pm$0.2 & 2.2$\pm$0.3$^{n}$ & 65.4$\pm$1.5$^{b}$ & 4.4$\pm$0.2 & 18.8$\pm$0.6 & \point\\
19B &0.894 &AGN& 4.6$\pm$0.6 &\point &\point &\point &\point &\point &\point \\
23A$^{\ast}$ &1.009&AGN& 5.8$\pm$0.4 &\point &\point &\point &\point &\point &\point \\ \noalign{\smallskip}
26A$^{\ast}$  &0.616 &AGN& 21.8$\pm$0.3 & 8.1$\pm$0.3 & 2.3$\pm$0.2$^{n}$ & 9.3$\pm$0.4$^{n}$ & 4.1$\pm$0.4 & 13.9$\pm$0.4 &\point \\
28B  &0.205&AGN&\point &\point &\point & 26.0$\pm$6.9$^{b}$ & 18.2$\pm$3.5 & 93.9$\pm$6.4 & 234.1$\pm$8.1$^{b}$ \\
29A &0.784&AGN& 9.0$\pm$0.4 & 6.2$\pm$0.5 & 5.9$\pm$0.5$^{n}$ &\point &\point &\point &\point \\
32A &1.113&AGN& 0.7$\pm$0.1 &\point &\point &\point &\point &\point &\point \\
37A &0.467&AGN& 9.4$\pm$0.4 &\point & 8.7$\pm$0.5$^{n}$ & 8.0$\pm$0.3$^{n}$ & 4.7$\pm$0.4 & 14.5$\pm$0.3 &\point \\ \noalign{\smallskip}
    & & & & &               & 31.6$\pm$1.1$^{b}$ &        &              & \\
45Z$^{\ast}$  &0.711&AGN& 22.0$\pm$1.0 & 7.2$\pm$0.8 &\point &\point &\point &\point &\point \\
48B$^{\ast}$ &0.498&AGN:& 16.0$\pm$3.0 &\point &\point &\point &\point &\point &\point \\
48Z &0.502 &/Grp:& 29.1$\pm$3.7 &\point &\point & 7.0$\pm$1.4$^{n}$ &\point & 6.5$\pm$1.3 &\point \\
51D &0.528&AGN&  4.7$\pm$0.3 &\point &\point &\point & 3.3$\pm$0.7 & 4.6$\pm$0.7 & \point\\ 
51L  &0.620&AGN& 13.0$\pm$0.6 & 2.2$\pm$0.3 & 2.4$\pm$0.3$^{n}$ & 6.1$\pm$0.4$^{n}$ & 6.8$\pm$0.5 & 19.7$\pm$0.6 & \point\\ \noalign{\smallskip}
53A  &0.245&GAL&\point &\point &\point &\point &\point & 5.0$\pm$0.6 & 6.8$\pm$0.6$^{n}$\\
56D &0.366&AGN & 2.8$\pm$0.2 & 2.1$\pm$0.2 &\point & 27.4$\pm$3.3$^{b}$ & 2.5$\pm$0.2 & 7.9$\pm$0.2 & \point\\
58B &0.629&Grp& 10.3$\pm$1.7 &\point &\point &\point &\point &\point &\point \\
59A &0.080 &AGN& 6.2$\pm$0.3 & 1.1$\pm$0.2 & \point& 1.2$\pm$0.1$^{n}$ & 2.7$\pm$0.2 & 2.6$\pm$0.2 & 11.2$\pm$0.1$^{n}$\\
     & & & & & & & & & 32.3$\pm$2.2$^{b}$ \\
61B  &0.592&AGN& 10.5$\pm$1.4 &\point &\point &\point &\point & 7.3$\pm$0.8 &\point \\ \noalign{\smallskip}
67A  &0.546&Grp& 10.1$\pm$0.9 &\point &\point & 1.0$\pm$0.4$^{m}$ &\point & 3.1$\pm$0.5 &\point \\
117Q &0.780&AGN& 46.0$\pm$2.3 & 13.2$\pm$1.6 &\point &\point &\point &\point &\point \\
      \noalign{\smallskip}
      \hline
      \end{tabular}
      \end{flushleft}
      \end{table*}

Three RDS quasars at redshifts larger than 2.4 have Ly${\alpha}$ emission lines in their optical spectra, one of them (54A) shows only a narrow component (EW$\sim$55 \AA). The objects 17A and 37G have broad Ly${\alpha}$ emission lines with EW of 55.8 \AA~and 36.2 \AA~(5.5 \AA -- narrow component), which do not differ much from the mean EW 52.7$\pm$3.5 \AA~of the GR sample, but are below the range of mean values of several other emission line and optical/radio quasar surveys (SSG 67--81 \AA).\\
\hspace*{5mm} Balmer lines: Two RDS AGNs 28B and 59A exhibit a broad H${\alpha}$ component (EW$=234.1$ \AA~and 32.3 \AA). The mean RIXOS H${\alpha}$ EW is 140$\pm$20 ($\sigma$$=$90) \AA, whereas the ST AGN sample has a mean H${\alpha}$ EW of 240$\pm$24 ($\sigma$$=$148) \AA. The object 59A shows in addition a narrow H${\alpha}$ emission line component (EW$=$11.2 \AA).
The mean H${\beta}$ (broad) EW of the RDS AGNs (38$\pm$8 \AA~with a dispersion $\sigma$$=$17 \AA) is consistent with the mean values of the RIXOS sample (38$\pm$3 \AA~with $\sigma$$=$23 \AA) and the CCRS sample (41$\pm$4 \AA~with $\sigma$$=$18 \AA), but about half of the values found from the ST AGN sample (77$\pm$6 \AA~with $\sigma$$=$45 \AA) and from BG AGN sample (96$\pm$4 \AA~with $\sigma$$=$37 \AA). The mean H${\beta}$ EW of the narrow component of the RDS AGNs (6$\pm$2 \AA~with $\sigma$$=$4 \AA) agrees quite well with the mean EW of 8$\pm$2 ($\sigma$$=$9) \AA, derived from 18 narrow emission lines of our field NELG sample. The mean H${\gamma}$ EW of the RDS AGNs is 4$\pm$1 ($\sigma$$=$3) \AA, where only narrow emission lines were detected. In the sample of 67 NELG we have found only five objects with H${\gamma}$ with a mean EW of 12$\pm$3 ($\sigma$$=$3) \AA.\\

%
   \begin{table*}[t]
      \caption{Distribution of the EW [\AA] of the RDS quasar lines.}
      \label{TAB10}
      \begin{flushleft}
      \begin{tabular}{lccccccccccc}
      \hline\noalign{\smallskip}
\hspace{0.5cm}lines   & \multicolumn{4}{c}{RDS} & RIXOS & CCRS &  GR & SS & SSG \\
  & mean &  $\sigma$ & error & n & mean & mean &mean &mean & range of means \\
\hspace{0.5cm}(1) & (2) & (3) & (4) & (5) & (6) & (7) & (8) & (9) & (10) \\ \noalign{\smallskip}
\hline \noalign{\smallskip}
SI~IV~$\lambda1397$          & 12      & 4  & 2  & 4  & -&- & 6$\pm$1 (57)&- &   -  \\
C~IV~$\lambda1548$           & 37~(54) & 20 & 6  & 9  & -&30$\pm$5 (15)&50$\pm5$ (63)&29$\pm$2 (51)& 26-60\\
C~III$]~\lambda1908$         & 41~(51) & 59 & 15 & 16 &22$\pm$3 (49)&18$\pm$2 (32)&16$\pm$2 (52)&18$\pm$1 (86)& 15-24\\
Mg~II~$\lambda2798$          & 46~(50) & 56 & 12 & 21 &37$\pm$3 (100)&29$\pm$2 (53)&- &35$\pm2$ (74)& 38-43\\ 
      \noalign{\smallskip}
      \hline
      \end{tabular}
      \end{flushleft}
      \end{table*}

%
%
   \begin{table*}[t]
      \caption{EW [\AA] distribution of RDS AGNs, of the comparison AGN samples and of the field NELG forbidden lines.}
      \label{TAB9}
      \begin{flushleft}
      \begin{tabular}{lccccccc@{\hspace*{6mm}}|@{\hspace*{6mm}}cccc}
      \hline\noalign{\smallskip}
\hspace{0.5cm}lines   & \multicolumn{4}{c}{RDS AGNs} & RIXOS & BG & ST &  \multicolumn{4}{c}{Field NELG}\\
           &  mean& $\sigma$ & error & n& mean & mean& mean & mean &  $\sigma$ & error & n\\
\hspace{0.5cm}(1) & (2) & (3) & (4) & (5) & (6) & (7) & (8) & (9) & (10) & (11) & (12)\\ \noalign{\smallskip}
\hline \noalign{\smallskip}
$[$Ne~V$]~\lambda3426$   & 4&  4& 1 & 9 &-&-& 7$\pm$2 (7) &- &- &- &  - \\
$[$O~II$]~\lambda3727$   & 12& 11& 3 &17 &-&-& 16$\pm$6 (15)&30&28& 4& 61\\
$[$Ne~III$]~\lambda3868$ &  5&  4& 2 & 8 &-&-& 10$\pm$2 (18) &4 & 2& 1& 8  \\
$[$O~III$]~\lambda4959$  & 6&  5& 2 & 8 &-&11$\pm$1 (70) & 14$\pm$3 (59) &8 & 8& 2& 21\\
$[$O~III$]~\lambda5007$  &20& 28& 9 & 9 &29$\pm$3 (59) &25$\pm$3 (83) &44$\pm$8 (62) &26&30& 6& 29\\
      \noalign{\smallskip}
      \hline
      \end{tabular}
      \end{flushleft}
      \end{table*}
%

Forbidden lines: In Table \ref{TAB9} we present the distribution of the rest frame EW of the RDS AGNs and of the AGN comparison samples as well of the field NLEG. The first column gives the name of the line. The columns 2--5 contain the mean EW, the standard deviation, the error on the mean and the number of emission lines for the RDS AGN sample. The columns 6--8 list for comparison the mean EW, its error and the number of objects of the RIXOS, the ST and the BR AGN sample. The columns 9--12 give the mean EW, the standard deviation, the error on the mean and the number of emission lines for our field NELG sample.

The mean EW of the forbidden lines of the RDS AGNs is slightly smaller than the values found for the AGN comparison samples, but they agree within the relatively large errors. The mean EW of the RDS AGNs seems to be in better agreement with the mean EW of the field NELG. Only the mean [O II] EW of the NELG is significantly larger than that 
obtained for the RDS AGNs and as well larger than derived for the comparison AGN samples. The mean [O~II] EW (30 \AA) of our NELG sample is well consistent with the average EW of $\sim33$ \AA~for CFRS galaxies (EW$>$15 \AA) with M$_{B}<-21$ mag and 0.4$<$z$<$0.75 (Hammer et al. \cite{Ham97}).  

Most of the AGNs (see Fig. \ref{spec1}), which exhibit narrow forbidden emission lines of oxygen and of neon, show a significant continuum originated in the host galaxy (see Sect. 3). This is probably the reason, that we find slightly smaller mean EW of the [Ne~III] and [O~III] RDS AGN emission lines in comparison with the other AGN samples. The mean [O II] EW should be less affected by the galaxy continuum located right of the line.

The ratio of mean [O~III] $\lambda5007$ to [O~II] $\lambda3727$ EW for the RDS AGNs ($\sim1.7$) is significantly higher than that ($\sim0.9$) found for the field NELG, which indicates a higher ionization parameter in AGNs compared to the field NELG.

In general the mean RDS AGN EW of the broad emission lines are consistent with the data from  several X--ray selected AGN samples at lower mean redshift (ST: $\left<z\right>$$=$0.2, RIXOS: $\left<z\right>$$=$0.8, CCRS: $\left<z\right>$$=$0.9) compared to its mean redshift of $\left<z\right>$$=$1.2. Further they are in general good agreement with the mean EW from UV/optical selected AGN samples at low and high redshift (BG: $\left<z\right>$$=$0.2, GR: $\left<z\right>$$=$0.5 and SS: $\left<z\right>$$=$1.6). 
Although our results are derived from a small number of emission lines they seem to 
indicate in comparison with the other samples nearly the same properties of AGN broad emission lines.

The mean EW of the RDS AGN narrow emission lines, found at redshifts below 1, are slightly smaller than found for the comparison AGN samples. There is a better agreement of the mean RDS AGN EW with those values obtained from our sample of narrow--emission line field galaxies (NELG). A reason for the less strong narrow emission lines of RDS AGNs compared to the other AGN samples could be a stronger contribution of the AGN host galaxies to the continuum emission (see discussion). 
 
\section{Fraction of NELG in field galaxies}

The comparison of the RDS with several deep ROSAT surveys (eg. CRSS, Boyle et al. \cite{Boy95a}; Deep ROSAT Survey (DRS), Georgantopoulous el al. \cite{Geor96}; UK ROSAT deep field survey (UKDS), McHardy et al. \cite{Mc98}) reveals that the RDS contains a larger percentage of AGNs than found in any previous survey (cf. Paper II).
This is partly due to the accurate HRI position of the X--ray sources, which allows a reliable optical identification and partly due to the high quality Keck spectra and our classification scheme, in which we have included the high ionization [Ne V] emission line as a sign for AGN activity. 
In contrast to the CCRS, the DRS and the UKDS, which have a fraction of 
narrow emission line galaxies (NELG) between 10 \% and 26 \%,
we did not find a significant number of NELG as X--ray counterparts, although some 
of our class d and e AGNs would have been classified as NELG in the previous X--ray surveys.\\
\hspace*{5mm} We have found only one NELG (53A) in our sample of 50 X--ray sources, which could still be argued against because of the confusion with two M--stars. 
Furthermore, the optical identification of the Ultra Deep HRI survey in the Lockman Hole (Hasinger et al. \cite{Has99b}) shows no evidence for a new population of faint galaxies, which has been suggested to dominate the soft X--ray source counts at X--ray fluxes down to 2.0$\cdot$10$^{-15}$ erg cm$^{-2}$ s$^{-1}$. \\
\hspace*{5mm} There are several hints that the identification of faint X--ray sources
with NELG is probably spurious due to relatively large PSPC error
circles and lower quality optical spectra. For example, five of 10 X--ray luminous NELG from the CCRS were identified as Seyfert galaxies (AGNs) by Boyle et al. (\cite{Boy95b}) using additional intermediate--resolution spectra. Nevertheless,
they argued that between 7 to 17 \% of the 2 keV X--ray background can be produced by starburst (NELG) galaxies.\\
\hspace*{5mm} We have available Keck spectra of 83 field galaxies near the positions of the X--ray sources. The spectra of these objects were either taken together with optical counterparts on long--slits, chosen as field objects on multi--slit masks or taken as possible counterparts of the X--ray sources. We have fitted a Gaussian to the emission lines of the field galaxies using the same algorithm as described in Sect. 5. We classify a field galaxy as a NELG, when the optical spectrum shows at least one narrow emission line (FWHM$<$1500 km s$^{-1}$) with a signal-to-noise ratio of S/N$\ge$3. The FWHM and the EW distributions of the NELG are already discussed in Sect. 5.1 and 5.2.                      .

\begin{figure}[t]
     \begin{minipage}{88mm}
     \psfig{figure=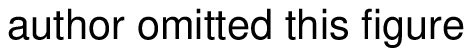,bbllx=71pt,bblly=255pt,bburx=323pt,bbury=444pt,clip=}
     \end{minipage}
     \caption[]{Keck LRIS spectra of four typical emission line galaxies (NELG)
                nearby the RDS sources. The wavelength is given in \AA ngstrom,
                the flux is in counts as shown in Fig. \ref{spec1}. 
                The positions of the two field objects
                14C and 26C are marked in the finding charts of the X-ray sources
                14 and 26.}
     \label{nelg}
\end{figure}

The sample of field objects contains 67 NELG (81 \%), 14 absorption line galaxies (17 \%) and 2 AGNs (2 \%). Fig.~\ref{nelg} presents the Keck spectra of four NELG, which show narrow Balmer emission lines (H${\alpha}$, H${\beta}$) and narrow forbidden lines [O~II]~$\lambda$3726, [O~III]~$\lambda$4959 and [O~III]~$\lambda$5007. Most NELG spectra contain at least a prominent [O~II] emission line. None of the spectra of the 67 NELG contains the high ionization narrow emission lines of [Ne V], which we have found in the spectra of RDS AGNs.

%
Guzman et al. (\cite{Guz97}) have done a spectroscopic study of 51 compact field galaxies in the flanking fields of the Hubble Deep Field, where about 88\% of the objects show narrow emission lines, and only absorption lines in the remaining 12\%. Therefore, both our sample and the Guzman sample of field galaxies suggest a high probability of misidentification of faint
X--ray sources with NELG in the field, if the presence of emission lines is used as the only indicator for identifications and the error boxes are larger than a few arcsec. The optical identification of all previous  deep X--ray surveys is based only on the PSPC detection. The RDS finding charts in Fig. \ref{spec1} show a number of cases (eg. 14Z, 117Q, 38A) where at least two faint NELG were within the PSPC error circle; if we had to rely on the PSPC observation alone, misidentification would have certainly arisen. 

\begin{figure}[t]
     \begin{minipage}{80mm}
        \psfig{figure=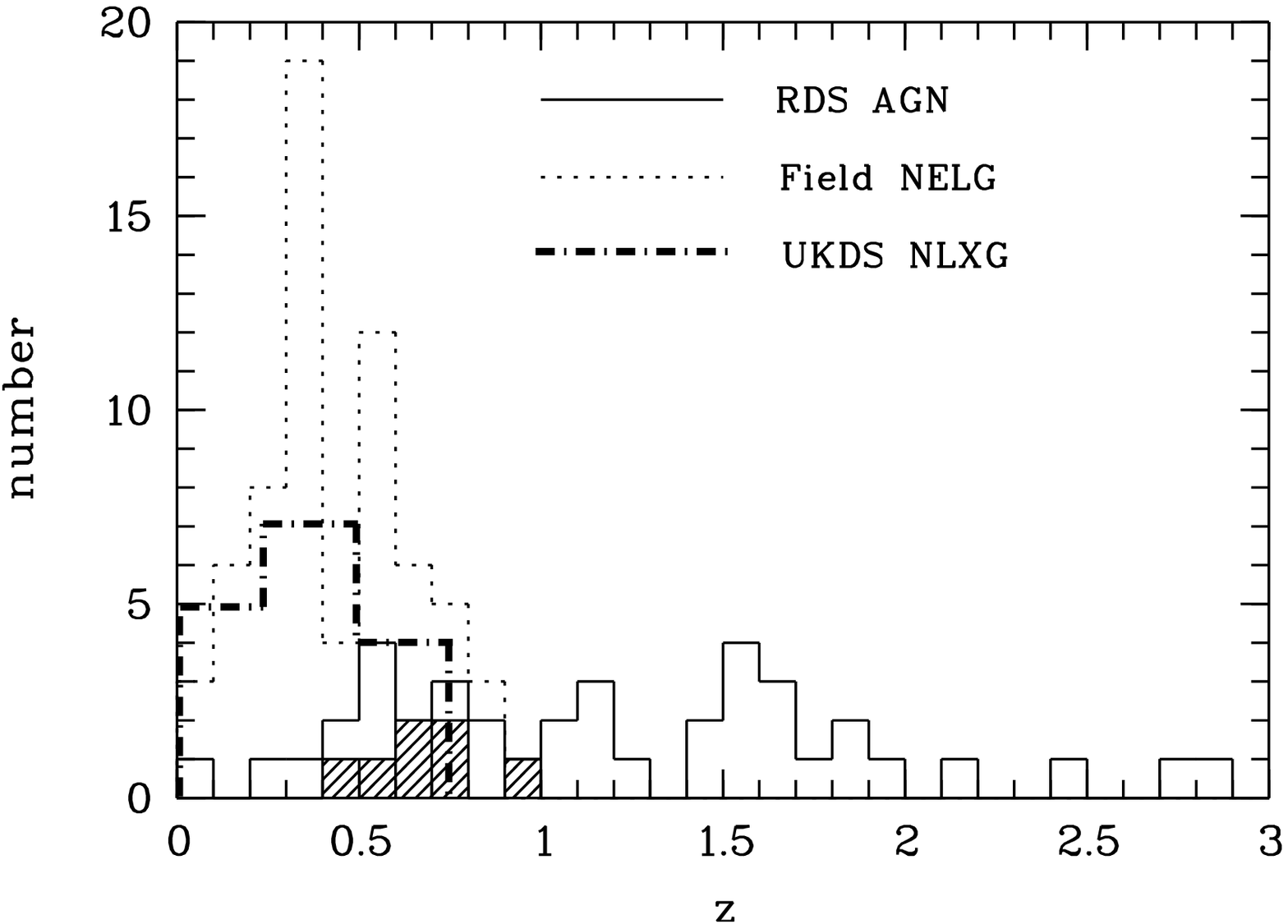,width=80mm,clip=}
     \end{minipage}
     \caption[]{The number-redshift relation for RDS AGNs (narrow line AGNs -- hatched histogram) and for field galaxies (NELG) close to the X--ray sources. The NELG distribution fits well with the redshift distribution of ''Narrow emission line X--ray galaxies'' (NLXG) of the UKDS.}
     \label{red}
\end{figure}

A further indication for a possible misidentification of faint X--ray sources 
comes from the redshift distribution of the RDS sources and the field NELG (see Fig. \ref{red}) in comparison with the distribution of NELG (called NLXG) in the UKDS where they make up a large fraction (26 \%). Jones et al. (\cite{Jon97}) have derived a double-peaked n($z$)
distribution of their UKDS objects, which peaks for the NLXG at $z\sim0.3$ and for the QSO at $z\sim1.6$. They argue that the difference in these
redshift distributions suggests two different source populations (QSO and NLXG), unless some NLXG could be AGNs (Seyfert II). If we compare the n($z$) distribution
of our field NELG with the UKDS NLXG, we find a good agreement, whereas the 
n($z$) of the RDS AGNs differs quite clearly (see Fig. \ref{red}). Also the 7 RDS narrow--line AGNs of ID classes d--e (see Table \ref{TAB1}) show a redshift distribution different from that of the UKDS NLXG.
This further suggests that a significant fraction of the UKDS sources is probably misidentified with field NELG.
%

%

\section{Discussion and conclusion}

We have presented the complete atlas of finding charts of 50 X--ray sources with 0.5--2.0
keV fluxes above 5.5 $\cdot$ 10$^{-15}$ erg cm$^{-2}$ s$^{-1}$ from the ROSAT Deep Survey 
in the Lockman Hole. For each of the 46 so far identified X--ray sources we have shown high quality spectra, most of them taken with the Keck telescopes.
One of four unidentified X--ray sources has been identified as an AGN through new spectroscopic Keck data. The existence of a double lobed radio galaxy detected with the VLA in the 6 cm and 20 cm band, leads to a tentative identification of a further previously unidentified source as a powerful radio galaxy (AGN). The two remaining optically unidentified X--ray sources show relatively bright counterparts in the $K^{\prime}$-band, which results in a very red colour of these sources ($R-K^{\prime}>5.0$).

There are several examples for faint X--ray sources identified with red objects (eg. Newsam et al. \cite{New97}). In the Ultra Deep HRI Survey with a 0.5--2.0 keV flux limit of 1.2 $\cdot$ 10$^{-15}$ erg cm$^{-2}$ s$^{-1}$ (Hasinger et al. \cite{Has99a}) we found that all X-ray counterparts with red colours ($R-K^{\prime}>4.5$) are either members of high redshift clusters of galaxies (z$>$1) or obscured AGNs at various redshifts.
Recently, we have detected a number of very red galaxies (Lehmann et al. \cite{Leh99}) indicating one of the most distant X--ray selected cluster of galaxies (Hasinger et al. \cite{Has99b}). The very red colour of the near--infrared counterparts of both RDS X--ray sources 14 and 84 suggests that they are obscured AGNs. Further observations are being performed to confirm these tentative identifications.

The quasar emission line properties of the RDS AGNs are in good agreement with the results from several X--ray selected and optical/UV selected AGN samples, while the mean EW of the [O~III], [Ne~III] and of the narrow components of the H${\gamma}$ and H${\beta}$ emission lines are slightly smaller than found for other AGN samples. This is probably due to a strong continuum flux originated in the AGN host galaxy. Many RDS AGNs show a significant galaxy continuum in their optical spectra as signified by their large D(4000) index values. The reason of a stronger contribution of the host galaxy in our AGNs could be a result of selection effects. The UV/optical selected AGN samples and shallower soft X--ray surveys may find AGNs, where the optical emission of the AGN is stronger than that of the host galaxies. In contrary, many of our RDS AGNs show a strong optical emission originated in the host galaxy in comparison with that from the AGN.

About 80 \% of the RDS AGNs show broad emission lines in their optical spectra (ID class a--c), whereas about 20 \% of our AGNs show only narrow emission lines (ID class d--e). Akiyama et al. (\cite{Aki99}) have obtained the same fraction ($\sim$ 20 \%) for the narrow--line AGNs from the optical identification of the ASCA LSS survey (Ueda et al. \cite{Ue98}), which has a flux limit of $\sim10^{-13}$ erg cm$^{-2}$ s$^{-1}$ in the 2--10 keV band. It is not possible to compare these numbers directly with the relative numbers of different AGN--types from the BeppoSAX HELLAS Survey in the 5--10 keV band (Fiore et al. \cite{Fio99}), because these authors have chosen a more complex diagnostic scheme. 

From our complete identification of the RDS sources and a comparison with a sample of field NELG we see no indication for a new class of X--ray bright narrow emission line
galaxies, which has been suggested to dominate the soft X--ray counts at fainter fluxes.

\begin{acknowledgements}
The ROSAT project is supported by the Bundesministerium
f\"ur Forschung und Technologie (BMFT), by the National
Aeronautics and Space Administration (NASA), and the Science
and Engineering Research Council (SERC). 

The W.~M. Keck Observatory is operated as a scientific partnership
between the California Institute of Technology, the University of
California, and the National Aeronautics and Space Administration. It
was made possible by the gerous financial support of the W.~M. Keck
Foundation. We thank the Institute for Astronony of the University
of Hawaii, Palomar Observatory, and the National Optical Astronomy
Observatories for grants of observing time.

I.L. and M.M. were visiting astronomer at the German-Spanish Astronomical Centre, Calar Alto, operated by the Max-Planck-Instiute for Astronomy, Heidelberg, jointly with the Spanish National Commission for Astronomy. I.L. acknowledges support from a DFG travel grant (HA 1850/11--1).
        
This work was supported
in part by NASA grants NAG5--1531 (M.S.), NAG8--794, NAG5--1649,
and NAGW--2508 (R.G.), and NSF grant AST--9509919 (D.S.). G.H. acknowledges the grant
FKZ 50 OR 9403 5 by the Deutsche Agentur f\"ur Raumfahrtangelegenheiten
(DARA). G.Z. acknowledges partial support by the Italian
Space Agency (ASI) under contract ARS--96--70 and the Italian Ministry
for University and Research (MURST) under grant Cofin 98--02--32.

We want to thank Axel Schwope for providing a number of MIDAS routines,
which were used to analyze the optical data.
\end{acknowledgements}

\end{document}